\documentclass[prd,amssymb,amsmath,amsfonts,superscriptaddress,nofootinbib,reprint,showpacs, longbibliography]{revtex4-1}

\usepackage{graphicx}
\usepackage{lmodern}
\usepackage{amsmath,amssymb}
\usepackage{mathrsfs}
\usepackage{amsfonts}
\usepackage[utf8]{inputenc}
\usepackage{url}
\usepackage[colorlinks]{hyperref}
\usepackage{xcolor}
\usepackage{multirow}
\usepackage[normalem]{ulem}
\usepackage{orcidlink}

\definecolor{dodgerblue}{HTML}{1E90FF}
\definecolor{viennared}{HTML}{DA0A14}
\definecolor{ctorange}{HTML}{FF6C0C}
\definecolor{wales}{HTML}{ff0038}
\definecolor{benettongreen}{HTML}{009421}
\definecolor{ferrarired}{HTML}{ff2800}
\definecolor{austriawienpurple}{HTML}{441678}
\definecolor{chitot}{RGB}{255,127,14}
\definecolor{chip}{RGB}{148,103,189}
\definecolor{shockingpink}{HTML}{FF00D4}
\definecolor{skyblue}{RGB}{37,121,217}

\hypersetup{
     colorlinks=true,
     linkcolor=dodgerblue,
     filecolor=dodgerblue,
     citecolor = dodgerblue,      
     urlcolor=dodgerblue,
     }

\newcommand{\Birmingham}{School of Physics and Astronomy and Institute for Gravitational Wave Astronomy, University of Birmingham, Edgbaston, Birmingham, B15 9TT, United Kingdom}

\newcommand{\nchip}{\vec{\chi}_\perp}

\newcommand{\tref}{t_{\rm ref}}

\begin{document}

\title{New effective precession spin for modeling multimodal gravitational waveforms in the strong-field regime}

\author{Lucy M. Thomas \orcidlink{0000-0003-3271-6436}}
\email{lthomas@star.sr.bham.ac.uk}
\affiliation{\Birmingham}
\author{Patricia Schmidt \orcidlink{0000-0003-1542-1791}}
\email{pschmidt@star.sr.bham.ac.uk}
\affiliation{\Birmingham}
\author{Geraint Pratten \orcidlink{0000-0003-4984-0775}}
\email{gpratten@star.sr.bham.ac.uk}
\affiliation{\Birmingham}

\date{\today}

\begin{abstract}
Accurately modeling the complete gravitational-wave signal from precessing binary black holes through the late inspiral, merger and ringdown remains a challenging problem. The lack of analytic solutions for the precession dynamics of generic double-spin systems, and the high dimensionality of the problem, obfuscate the incorporation of strong-field spin-precession information into semianalytic waveform models used in gravitational-wave data analysis. Previously, an effective precession spin $\chi_p$ was introduced to reduce the number of spin degrees of freedom. Here, we show that $\chi_p$ alone does not accurately reproduce higher-order multipolar modes, in particular the ones that carry strong imprints due to precession such as the $(2,1)$-mode. 
To improve the higher-mode content, and in particular to facilitate an accurate incorporation of precession effects in the strong-field regime into waveform models, we introduce a new dimensional reduction through an \emph{effective precession spin vector}, $\nchip$, which takes into account precessing spin information from both black holes. We show that this adapted effective precession spin (i) mimics the precession dynamics of the fully precessing configuration remarkably well, (ii) captures the signature features of precession in higher-order modes, and (iii) reproduces the final state of the remnant black hole with high accuracy for the overwhelming majority of configurations. We demonstrate the efficacy of this two-dimensional precession spin in the strong-field regime, paving the path for meaningful calibration of the precessing sector of semianalytic waveform models with a faithful representation of higher-order modes through merger and the remnant black hole spin.
\end{abstract}

\maketitle 

\section{Introduction}
\label{sec:intro}
Since the first detection of gravitational waves (GWs) from a binary black hole (BBH) merger in 2015 \cite{Abbott:2016blz}, the GW observatories Advanced LIGO~\cite{Aasi:2013wya,Tse:2019wcy} and Virgo~\cite{TheVirgo:2014hva,Acernese:2019sbr} have reported detections of GWs from tens of compact binary mergers~\cite{LIGOScientific:2018mvr, GWTC2}, including the first multimessenger observation of a binary neutron star inspiral, GW170817~\cite{GW170817-DETECTION, GW170817-MMA}, the first intermediate mass black hole GW190521~\cite{GW190521-DETECTION, GW190521-ASTRO}, the first unequal-mass BBH GW190412~\cite{GW190412-DETECTION} and the first neutron star--black hole candidates GW190814~\cite{GW190814-DETECTION} and GW190426~\cite{GWTC2}. Additional GW candidates have been reported from analyses of the publicly available data in Refs.~\cite{Nitz:2019hdf, Venumadhav:2019lyq}.
In order to fully exploit the scientific potential of these observations and infer the source properties, highly accurate and computationally efficient models of the emitted GW signal are required. 

Recent years have seen significant improvements in the modeling of the complete inspiral-merger-ringdown (IMR) signal of compact binaries with the inclusion of spin-induced precession effects~\cite{Hannam:2013oca, Pan:2013rra} as well as higher-order harmonics~\cite{Khan:2019kot, Pratten:2020ceb, Ossokine:2020kjp}. 
While the state-of-the-art waveform models are sufficiently accurate for current observations, where the uncertainty in the measurement of the BH properties is dominated by the statistical uncertainty due to detector noise, future upgrades to the current interferometer network~\cite{Reitze:2019iox} and third-generation ground-based detectors such as the Einstein Telescope \cite{Maggiore:2019uih} and Cosmic Explorer~\cite{Evans:2016mbw}, will operate at unprecedented sensitivities, shifting focus onto systematic modeling errors as the dominant source of error~\cite{Purrer:2019jcp}. The development of evermore accurate models by increasing their physics content is of paramount importance.

The current generation of waveform models can broadly be split into three families: effective-one-body (EOB) models~\cite{Buonanno:1998gg}, phenomenological (Phenom) models~\cite{Ajith:2009bn, Santamaria:2010yb}, and numerical relativity (NR) surrogate models~\cite{Blackman:2017dfb, Blackman:2017pcm}. While the EOB and Phenom models describe the complete GW signal throughout inspiral, merger and ringdown, NR surrogates are mainly restricted to the strong-field regime. Phenom models are currently the most widely used due to their considerable computational efficiency relative to other models, which is particularly crucial for parameter estimation due to the large number of required model evaluations, and their wide parameter space validity. Computational efficiency, however, often comes at the cost of simplified physics content, which can severely impact the accuracy of parameter measurements~\cite{Ramos-Buades:2020noq}. In this paper, we focus on one particularly urgent open problem in waveform modeling: devising a feasible strategy to accurately incorporate precession effects into BBH waveform models in the strong-field regime. 

Spin-induced orbital precession occurs when the spins of one or both compact objects are misaligned with the orbital angular momentum~\cite{Apostolatos:1994mx, Kidder:1995zr}. This introduces phase and amplitude modulations into the GW signal, and a richer mode structure, amplifying higher-order modes (HMs) relative to the $(2,2)$-mode. Due to their complexity, precessing waveforms encode vast amounts of information which can be used to break parameter degeneracies~\cite{Vecchio:2003tn, vanderSluys:2007st, Lang:1900bz, Cho:2012ed, OShaughnessy:2014shr, Chatziioannou:2014coa, Pratten:2020igi}. This facilitates better measurements and more stringent tests of general relativity, but this also makes them difficult to model across the binary parameter space. 

Current precessing IMR waveform models are built in an approximate way by applying a time-dependent rotation encoding the orbital precession dynamics to waveform modes obtained in a frame that coprecesses with the orbit~\cite{Schmidt:2010it, Schmidt:2012rh}. 
The four spin components instantaneously orthogonal to the orbital angular momentum, i.e. within the instantaneous orbit plane, source the orbital precession~\cite{Apostolatos:1994mx}. This relatively large number of spin degrees of freedoms complicates the inclusion of precession into semianalytic waveform models. Therefore, developing an efficient dimensional reduction strategy is crucial for modeling strong-field precession across the parameter space.

Previously, the effective precession spin parameter $\chi_p$~\cite{Schmidt:2014iyl} was introduced to reduce the four orthogonal spin components to one while capturing the dominant precession effects in the waveforms. However, a simple $\chi_p$-parametrization fails to accurately reproduce the phenomenology of HMs, including precession-induced mode mixing and the asymmetry between positive and negative $m$-modes \cite{Ramos-Buades:2020noq}. In precessing systems where the relative power in HMs can be comparable to the dominant quadrupolar mode, this can lead to significant systematic errors~\cite{Pratten:2020ceb, Ossokine:2020kjp}, as recent observations are starting to indicate~\cite{GW190521-ASTRO}. 

Here, we introduce a new two-dimensional effective precession spin vector, $\nchip$, which incorporates two-spin effects. Focusing on the strong-field regime, we show that a $\nchip$-parametrization, (i) matches the opening angle of the precession cone at a given reference time; (ii) significantly better reproduces HMs than a $\chi_p$-parametrization; (iii) more accurately mimics the precession dynamics, and (iv) matches the final state.
This vectorial effective spin mapping could facilitate more accurate waveform modeling of precession in the strong-field regime.

The paper is organized as follows: In Sec.~\ref{sec:Phenom} we briefly summarize the phenomenology of precessing binaries and current waveform modeling efforts, before introducing the new effective precession spin vector $\nchip$ in Sec.~\ref{sec:newmap}. We describe the methodology used in this work
in Sec.~\ref{sec:Methods}. In Sec. \ref{sec:Results} we present our results and subsequently discuss the accuracy and caveats of this spin mapping in Sec.~\ref{sec:Discussion}. Throughout this paper, we use $G=c=1$.

\section{Precessing Binaries}
\label{sec:Prec}

\subsection{Modeling precession}
\label{sec:Phenom}
Binary black holes on quasispherical orbits are intrinsically characterized by seven parameters: the mass ratio $q = m_1/m_2 \geq 1$, where $m_i$ with $i\in [1,2]$ denotes the component mass, and the spin angular momentum $\vec{S}_i$ of each black hole, or its dimensionless counterpart $\vec{\chi}_i = \vec{S}_i/m_i^2$.
If the black holes' spins are misaligned with the direction of the orbital angular momentum $\hat{L}$\footnote{We note the difference between the orbital angular momentum $\hat{L}$ and its Newtonian approximation $\hat{L}_{\rm N}$; however, in this work we will not distinguish between them.} of the binary motion, spin-induced precession of the orbital plane occurs~\cite{Apostolatos:1994mx, Kidder:1995zr}. This introduces characteristic amplitude and phase modulations in the GW signal, excites HMs, and modifies the final state of the merger remnant. 
The precession of the orbital plane is driven by the spin components $\vec{S}_{1\perp}$ and $\vec{S}_{2\perp}$ instantaneously perpendicular to $\hat{L}$, defined as $\vec{S}_{i\perp}=\vec{S}_i \times \hat{L}$. 
In a precessing binary, $\hat{L}$ and the orientation of the two spins $\hat{S}_i$ become time dependent. In the case of simple precession, throughout the inspiral, $\hat{L}(t)$ traces a cone centered around the direction of the total angular momentum $\hat{J}$, which remains approximately spatially fixed \cite{Apostolatos:1994mx, Kidder:1995zr}, i.e., $\hat{J}(t) \simeq \hat{J}_{t \rightarrow -\infty} \quad \forall t$, where $\vec{J} = \vec{L} + \vec{S}_1 + \vec{S}_2$. The opening angle of this precession cone $\lambda_L(t)$ is defined as~\cite{Apostolatos:1994mx}
\begin{equation}
    \cos(\lambda_L(t)) \equiv \hat{L}(t) \cdot \hat{J}(t) = \frac{L(t)+S_\parallel(t)}{\sqrt{(L(t)+S_\parallel(t))^2+S_\perp^2(t)}},
\label{eq:cone}
\end{equation}
where $\vec{S}(t) = \vec{S}_1(t) + \vec{S}_2(t)$ is the total spin of the binary with $S_\perp(t) = ||\vec{S}(t) \times \hat{L}(t) ||$ and $S_\parallel(t) = \vec{S}(t) \cdot \hat{L}(t)$. 
The precession cone opening angle grows on the precession timescale, which lies between the shorter orbital timescale and the longer radiation reaction timescale, i.e., the time it takes for the binary to merge. This separation of timescales allows us to define approximate closed-form solutions of the post-Newtonian (PN) precession equations in the inspiral~\cite{Kesden:2014sla, Chatziioannou:2014coa, Gerosa:2015tea}. Due to the high dimensionality of the problem, however, precessing waveforms that include inspiral, merger, and ringdown are commonly modeled by applying a time-dependent rotation $\mathbf{R}(t)$ to the waveform modes obtained in a coprecessing frame that tracks the precession of the orbital plane~\cite{Schmidt:2010it, Schmidt:2012rh}, i.e.,
\begin{equation}
    h^P_{\ell m}(t) \simeq \sum_{m'=-\ell}^\ell \mathbf{R}_{\ell m m'}(t) h^\mathrm{co-prec}_{\ell m'}(t).
    \label{eq:hP}
\end{equation}
Such a decomposition is possible due to the approximate decoupling between the precession and inspiral dynamics~\cite{Schmidt:2012rh}. In addition, to simplify the problem, the coprecessing frame modes may be identified with aligned-spin modes~\cite{Schmidt:2012rh, Pekowsky:2013ska} as is done for the current generation of phenomenological waveform models~\cite{Hannam:2013oca, Khan:2019kot, Pratten:2020ceb}. We note, however, that this approximation introduces significant systematic errors due to the neglect of spin-induced mode asymmetries in precessing systems~\cite{Ramos-Buades:2020noq}. 
Additionally, no semianalytic precessing IMR waveform model currently incorporates NR information in modeling the precession dynamics through merger, again due to the high dimensionality of the problem. Efficient dimensional reduction strategies such as the use of \emph{effective parametrizations} to reduce the number of spin degrees of freedom may be a way forward to calibrate the precession dynamics in the strong-field regime. Insights from PN theory have previously led to the construction of an effective precession spin $\chi_p$ defined as \cite{Schmidt:2014iyl}
\begin{align}
    S_p &:= \max (A_1 S_{1\perp}, A_2 S_{2\perp}),\label{eq:s_p def}\\
    \chi_p &:= \frac{S_p}{A_1 m_1^2}, \label{eq:chi_p def}
\end{align}
where $A_1 = 2 + 3/2q$, $A_2 = 2 + 3q/2$, and $S_{i\perp}=||\vec{S}_{i\perp}||$ such that the Kerr limit $\chi_i \leq 1$ is obeyed. It is constructed such that it captures the average amount of precession exhibited by a generically precessing system over many precession cycles defined at some reference time $\tref$ during the inspiral. We note that $\chi_p$ will assume a (slightly) different value depending on $\tref$ but this time (frequency) dependence can be mitigated through the inclusion of additional precession-averaged spin effects~\cite{Gerosa:2020aiw}. An alternative effective parametrization based on the total spin can be found in~\cite{Akcay:2020qrj}.

The effective precession spin $\chi_p$ is regularly used to make statements about the measurement of precession at a certain reference frequency (time) in GW inference, see e.g.~\cite{TheLIGOScientific:2016wfe, LIGOScientific:2018mvr, Pratten:2020igi}, and may also present a natural way for calibrating precession effects in the strong-field through a single scalar parameter via the following effective mapping at some reference time $\tref$:
\begin{align}
    \vec{\chi}_1(\tref) &= (\chi_{1x}, \chi_{1y}, \chi_{1z}) \mapsto \vec{\chi}'_1 = (\chi_p, 0, \chi_{1z}), \\[8pt]
    \vec{\chi}_2(\tref) &= (\chi_{2x}, \chi_{2y}, \chi_{2z}) \mapsto \vec{\chi}'_2 = (0,0,\chi_{2z}),
    \label{eq:chipmap}
\end{align}
where the spin components are defined in a Cartesian binary source frame with $\hat{L}(\tref)=\hat{z}$. Such an identification reduces the four in-plane spin components to a single scalar quantity, making the problem of incorporating precession effects more tractable. This approach has successfully been implemented in the widely used phenomenological waveform approximant \texttt{IMRPhenomPv2}~\cite{Hannam:2013oca}.

The efficacy of such a $\chi_p$-parametrization, however, has only been demonstrated in the inspiral~\cite{Schmidt:2014iyl} focusing on the $(2,2)$-mode. HMs, however, are particularly important in binaries with large mass and spin asymmetries, for which also precession effects are more pronounced. 
While the radiation from a nonprecessing binary is dominated by the quadrupolar $(2,2)$-mode, which is predominantly emitted along $\hat{L}$, in a precessing system power is transferred from the $(2,2)$-mode to HMs.
These HMs can become comparable in strength to the quadrupolar mode in the later inspiral and merger, and some modes, especially the $(2,\pm1)$-modes, can be particularly strong~\cite{Schmidt:2010it}. 
Therefore, the accurate modeling of HMs is particularly important in precessing systems. We will show in Sec.~\ref{sec:modematches} that the simple $\chi_p$-parametrization of Eqs.~\eqref{eq:chipmap} fails to accurately reproduce the behavior of precessing HMs, motivating the introduction of a new effective precession spin vector $\nchip$ to address this issue.

\subsection{A new effective precession spin}
\label{sec:newmap}
To aid the calibration of the precessing sector of semianalytic IMR waveform models, we seek to capture the dominant behavior through dimensional reduction by reducing the number of in-plane spin components through an effective map. To this end, we introduce a new dimensionless \emph{effective precession spin vector}, $\nchip(t) \in \mathbb{R}^2$. 

Our starting point for the construction of $\nchip$ is the opening angle of the precession cone at a reference time $t=\tref$, $\lambda_L(\tref)$ given by Eq.\eqref{eq:cone}, which captures the amount of precession in the system. We recall that the opening angle depends explicitly on the in-plane spin components through $S_\perp (t)$; we therefore seek a mapping such that this quantity is approximately preserved at the reference time at which the mapping is applied. To do so, we first place the in-plane spin projection of the total spin of the system onto the larger black hole, such that
\begin{equation}
    \vec{\chi}_{1\perp}(\tref) \mapsto \vec{S}_\perp(\tref) / m_1^2, \quad \vec{\chi}_{2\perp}(\tref) \mapsto \vec{0},
\end{equation}
where 
\begin{equation}
\vec{S}_\perp(\tref) = m_1^2 \, \vec{\chi}_{1\perp}(\tref) + m_2^2 \, \vec{\chi}_{2\perp}(\tref).
\end{equation}
We find, however, that this mapping can be further improved by assigning it conditionally to either the primary or secondary BH, depending on which BH has the largest in-plane spin magnitude $S_{i\perp}(\tref)$ at the reference time.
This conditional placement ensures that a binary with an in-plane spin on only one BH is correctly reproduced. Furthermore, we impose the Kerr limit on the BH spin by including appropriate normalization factors into the definition of $\nchip$. With these constraints, we obtain the following effective precession spin vector $\nchip(\tref)$:
\begin{align}
    \label{eq:ChiHatNorm}
    \nchip (\tref)&\equiv
    \begin{cases}
    \dfrac{\vec{S}_\perp}{m_1^2+ S_{2\perp}},& \text{if } S_{1 \perp} \geq S_{2 \perp}, \\[8pt]
    \dfrac{\vec{S}_\perp}{m_2^2 + S_{1\perp}},& \text{if } S_{1 \perp} < S_{2 \perp},
    \end{cases}
\end{align}
where the quantities $S_{1\perp}$, $S_{2\perp}$, $\vec{S}_{\perp}$ are all evaluated at $t_{\text{ref}}$.
We stress that the mass ratio $q$ and the spin components along $\hat{L}(\tref)$ remain unaltered in this particular mapping. Explicitly, in a Cartesian binary source frame with $\hat{L}(\tref) \equiv \hat{z}$, we have
\begin{align}
    \vec{\chi}_1&=(\chi_{1x}, \chi_{1y}, \chi_{2z}) \mapsto \vec{\chi}'_1=(\chi_{\perp x}, \chi_{\perp y}, \chi_{1z}) \\[8pt]
    \vec{\chi}_2&=(\chi_{2x}, \chi_{2y}, \chi_{2z}) \mapsto \vec{\chi}'_2=(0, 0, \chi_{2z}),
\end{align}
for $S_{1\perp}(\tref) \geq S_{2\perp}(\tref)$ and $1 \leftrightarrow 2$ else.
We note that instead of Cartesian coordinates, polar coordinates may be chosen. Then, $||\nchip||$ represents the magnitude of the mapped dimensionless spin vector and the azimuthal orientation, $\phi_\perp$, is its angular position within the orbital plane at the reference time.
We demonstrate the efficacy of this vectorial parametrization, in particular for HMs, in Sec.~\ref{sec:Results}.

\section{Methodology}
\label{sec:Methods}
\subsection{Waveforms}
\label{sec:wavegen}
To assess the efficacy of the new effective parametrization Eq.~\eqref{eq:ChiHatNorm}, we compare the waveforms obtained from the seven-dimensional system characterized by $(q, \vec{\chi}_1, \vec{\chi}_2)$ to the five-dimensional effective system described by $(q, \chi_{1\parallel}, \chi_{2\parallel}, \nchip)$ as well as to the four-dimensional system given by $(q, \chi_{1\parallel}, \chi_{2\parallel}, \chi_p)$. 
As we are particularly interested in testing the efficacy of such mappings in the strong-field regime, we use the NR surrogate model \texttt{NRSur7dq4}~\cite{Varma:2019csw} as provided through the public \texttt{GWSURROGATE} \textsc{PYTHON} package~\cite{gwsurrogate} to generate late inspiral-merger-ringdown waveforms for all our analyses.
The computational efficiency of this model allows us to assess the mappings over a dense sampling of the intrinsic parameter space. However, due to the limited parameter ranges of the NR simulations it is built upon, the surrogate is limited to dimensionless spin magnitudes $||\chi_i||\leq 0.8$ and mass ratios $q\leq4$. While precession effects are even more pronounced for higher mass ratios, the importance of the in-plane spin on the smaller BH decreases and therefore we expect any dimensional reduction that is built to capture the dominant precession spin to perform even better in this limit. 

The surrogate model represents an interpolant across a discrete set of NR simulations~\cite{Field:2013cfa, Blackman:2015pia, Blackman:2017dfb}. The precessing waveform modes up to $\ell \leq 4$ are obtained by following the strategy outlined in Sec.~\ref{sec:Prec}, where the coprecessing modes are further decomposed into co-orbital modes to further simplify their structure,
\begin{equation}
    h_{\ell m}^{\text{co-prec}}(t)=e^{i m \Omega(t)}h_{\ell m}^{\text{coorb}}(t),
\end{equation}
where $\Omega(t)$ is the relative angular velocity relating the two frames~\cite{Blackman:2017pcm}.

Unlike most other waveform models, the surrogate makes use of four unit quaternion components $\{\hat{q}_0(t), \hat{q}_1(t), \hat{q}_2(t), \hat{q}_3(t)\}$ instead of three Euler angles to describe the precession dynamics of the orbital plane~\cite{Boyle:2013nka}. Importantly, the precessing modes are obtained in an inertial frame corresponding to $\hat{L}(t_0) \equiv \hat{z}$ at the initial time $t_0$, as opposed to the more commonly used $\hat{J}$-aligned frame. In this coordinate frame, the $xy$-plane coincides with the initial orbital plane of the binary with the $x$-axis parallel to the separation vector pointing from the smaller black hole to the larger one. Due to this binary source frame choice, caution must be taken when interpreting the physical meaning of the quaternions.

The resulting waveforms are of a fixed length, from $t_0=-4300M$, the negative sign indicating premerger, up to $t=+100M$ after the merger. This relatively short length makes them unsuitable for describing the waveforms of low mass binaries $M\lesssim 70M_{\odot}$ assuming a starting frequency of 20 Hz. The surrogate determines the coalescence time $t_c$ as the peak of the quadrature sum of the mode amplitudes, 
\begin{equation}
\label{eq:peak}
    t_c = \max_t \sqrt{\sum_{\ell m}\left|h_{\ell m}(t)\right|^2},
\end{equation}
and shifts the time arrays such that the peak amplitude occurs at $t_c=0$. 

\subsection{Faithfulness for precessing waveforms}
\label{sec:MatchFormulae}
For our quantitative comparisons, we define $h$ to represent the fully precessing waveform with all 6 spin degrees of freedom, and $h_{\sigma}$ the corresponding waveform produced by an effective mapping, where $\sigma\in\left[\chi_p,\nchip\right]$. Hereafter, we refer to $h$ as the signal waveform, and to $h_{\sigma}$ as the template waveform. The effective mappings are applied at the surrogate initial time $t_0$, such that the full and mapped spins are used as initial data to produce the signal and template waveforms respectively.
To quantify how well either mapping reproduces the full waveform, we compute the match (faithfulness) between $h$ and $h_\sigma$, which is defined as the noise-weighted inner product between the two waveforms maximized over a time and phase shift of the template waveform:
\begin{equation}
    \label{eq:Match}
    \mathcal{M}(h,h_\sigma) = \max_{t_{c\sigma}, \phi_{0\sigma}}\frac{\langle h, h_\sigma\rangle}{\sqrt{\langle h, h\rangle \langle h_\sigma, h_\sigma\rangle}},
\end{equation}
where the inner product is defined as
\begin{equation}
    \label{eq:InnerProduct}
    \langle h,h_\sigma \rangle = 4\Re \int^{f_{\rm{max}}}_{f_{\rm{min}}} \frac{\tilde{h}(f)\tilde{h}_{\sigma}^*(f)}{S_n (f)},
\end{equation}
with $S_n(f)$ the one-sided power spectral density (PSD) of the detector noise, $\tilde{h}$ indicates the Fourier transform of $h$, and ``${}^*$'' complex conjugation.
In what follows, $h$ and $h_\sigma$ either denote individual waveform modes $h_{\ell m}$, or the complex strain defined as
\begin{equation}
    \label{eq:ModesDef}
    h\left(t,\theta, \phi\right) = \sum_{\ell}\sum_{m=-\ell}^{m=\ell} h_{\ell m}(t) {}^{-2}Y_{\ell m}\left(\theta, \phi \right),
\end{equation}
where ${}^{-2}Y_{\ell m}(\theta, \phi)$ are the spin-weighted spherical harmonics of spin weight $s=-2$ and $(\theta, \phi)$ are the polar and azimuthal angles on the unit sphere in the binary source frame. 

To assess how accurately individual modes, in particular HMs, are reproduced under the effective mapping, we compute individual mode-by-mode matches between each spin mapping and the full-spin waveform; i.e. for each pair $(\ell, m)$, $h$, $h_{\sigma}$ in Eq.~\eqref{eq:Match} are replaced by individual modes $h\rightarrow h_{\ell m}$ and $h_{\sigma}\rightarrow h_{\sigma, \ell m}$. 

As the odd $m$-modes are sourced by mass and spin asymmetries,
they are often contaminated by numerical noise for systems with small asymmetries. We therefore employ an additional cut on the energy $E_{\ell m}$ contained in the inertial-frame $(2,\pm1)-$ and $(3,\pm3)$-modes of the fully spinning mode prior to calculating the match, where the mode energy is given by
\begin{equation}
    E_{\ell m} = \frac{1}{16 \pi} \int_{t_0}^{t_f} |\dot{h}_{\ell m} (\tau)|d\tau,
\end{equation}
where $t_f$ is the final time of the surrogate waveforms. Based on calculations of the energy contained in those modes for binaries without mass or spin asymmetries, we find the energy thresholds for these modes given by the values listed in Table~\ref{tab:EnergyinModes}. Modes with $E_{\ell m}$ less than these values are discarded in the mode-by-mode match calculations performed in Sec.~\ref{sec:modematches}.

\begin{table}[t!]
    \centering
    \begin{tabular}{l|c}
    \hline
         ($\ell$,m)-mode & $E_{\ell m}$ threshold  \\
         \hline
         \hline
         $(2,\pm1)$ & $1.0\times 10^{-6}$ \\
         $(3,\pm3)$ & $5.5\times10^{-7}$ \\
         \hline 
    \end{tabular}
    \caption{Mode energy thresholds for odd $m$-modes. If the energy of a particular mode is below its threshold, the mode is considered to be numerical noise and excluded from the mode-by-mode match calculation.}
    \label{tab:EnergyinModes}
\end{table}

We perform mode-by-mode match calculations using both white noise, i.e. $S_n(f) =1$ and the projected aLIGO PSD for the fourth observing run \cite{NoiseCurves}, denoted $\mathcal{M}_{\text{white}}$ and $\mathcal{M}_{\text{O4}}$ respectively. 
The white noise matches are to assess the systematic errors induced by the mappings in the absence of detector-specific frequency sensitivities, while the PSD-weighted matches demonstrate the effect for a given detector.
For the detector PSD matches we choose a starting frequency of $f_{\rm{min}}=20$Hz, and truncate the waveforms at $t=50M$ after the peak as determined by Eq.~\eqref{eq:peak} to remove postmerger numerical noise. 

While the individual mode matches allow us to assess how well HMs in particular are captured by the lower-dimensional spin parametrization, GW detectors measure the strain, which also depends on extrinsic parameters of the source such as the luminosity distance, the effective polarization angle $\kappa$~\cite{Capano:2013raa} and the binary inclination $\iota$ relative to the line of sight of an observer.

Following Refs.~\cite{Babak:2016tgq,Ossokine:2020kjp, Pratten:2020igi}, we compute the strain match by analytically optimizing over the template polarization angle $\kappa_\sigma$ and numerically optimizing over the template reference phase $\phi_{0\sigma}$ and template coalescence time $t_{c\sigma}$,
\begin{equation}
\label{eq:strainmatch}
    \mathcal{M}_{\rm strain}(M, \iota, \phi_0, \kappa) = \max_{t_{c\sigma}, \phi_{0\sigma}, \kappa_\sigma} \frac{\langle h, h_\sigma \rangle}{\sqrt{\langle h, h \rangle \langle h_\sigma, h_\sigma \rangle}}\Bigg|_{\iota=\iota_\sigma}.
\end{equation}
We do not optimize over any intrinsic parameters. We note that Eq.~\eqref{eq:strainmatch} still depends on the signal polarization $\kappa$ and reference phase $\phi_0$. By averaging over these two angles, we obtain the sky-and-polarization-averaged strain match,
\begin{equation}
\label{eq:SkyPolAveragedStrainMatch}
    \overline{\mathcal{M}}_{\rm strain} (M,\iota)= \frac{1}{8\pi^2}\int^{2\pi}_{0}d\kappa \int^{2\pi}_{0} d\phi_0 \mathcal{M}_{\rm strain}(M,\iota,\phi_0,\kappa).
\end{equation}

Additionally, to account for the correlation between low matches and low signal-to-noise ratio (SNR), we also compute the SNR-weighted strain match~\cite{Ossokine:2020kjp, Pratten:2020igi} given by
\begin{equation}
    \label{eq:SNRWeightedStrainMatch}
    \overline{\mathcal{M}}_{\rm SNR}(M,\iota)= \left(\frac{\sum_{i} (\mathcal{M}(h, h_\sigma))^3 \langle h_i,h_i\rangle^{3/2}}{\sum_{i}\langle h_i,h_i\rangle^{3/2}}\right)^{1/3},
\end{equation}
where the sum is over a discrete range of source polarizations $\kappa$ and initial phases $\phi_0$ as detailed in Sec.~\ref{sec:Sampling}. 

We note that we do not apply the postmerger truncation at $t=50M$, nor do we impose the mode energy thresholds of Table~\ref{tab:EnergyinModes} when computing the sky-and-polarization-averaged and the SNR-weighted strain matches.
For strain matches we take into account all modes up to $\ell=4$ as provided by the NR surrogate.

Lastly, rather than showing the agreement between two waveforms, it can be advantageous to quantify the disagreement through the mismatch $\mathcal{MM}$ instead:
\begin{align}
    \label{eq:StrainMismatch}
    \overline{\mathcal{M}\mathcal{M}}_{\rm strain} &= 1 - \overline{\mathcal{M}}_{\rm strain},\\
    \label{eq:SNRMismatch}
    \overline{\mathcal{M}\mathcal{M}}_{\rm SNR} &= 1 - \overline{\mathcal{M}}_{\rm SNR}.
\end{align}

\subsection{Binary configurations}
\label{sec:Sampling}
\begin{table*}[t!]
    \centering
    \begin{tabular}{l|c|c|c}
    \hline
    \hline
    & Mode-by-mode matches & \begin{tabular}{@{}c@{}}Sky-and-polarization- \\averaged strain matches \end{tabular}& SNR-weighted strain matches \\
    \hline
         Spin magnitudes & \begin{tabular}{@{}c@{}c@{}}$||\vec{\chi}_1||$, $||\vec{\chi}_2||\in [0,0.1,0.2,0.3,0.4,$\\ $0.5,0.6,0.7,0.8]$, \\ if $||\vec{\chi}_1||$ =0, $||\vec{\chi}_2|| \neq 0$\end{tabular} & $||\vec{\chi}_1||$, $||\vec{\chi}_2|| \in U [0,0.8]$ &  $||\vec{\chi}_1||$, $||\vec{\chi}_2|| \in U[0,0.8]$\\
    \hline
         Tilt angles (rad) & \begin{tabular}{@{}c@{}}$\theta_1,\theta_2 \in [0, \pi/3, 2\pi/3, \pi]$, \\ if $\theta_1\in [0,\pi]$, $\theta_2\notin [0,\pi]$\end{tabular} & $\cos(\theta_1),\cos(\theta_2) \in U [-1,1]$ & $\cos(\theta_1),\cos(\theta_2) \in U [-1,1]$\\
    \hline
         Azimuthal angles (rad) & \begin{tabular}{@{}c@{}} $\phi_1=\pi/3$,\\ $\phi_2 \in [0, \pi/3, 2\pi/3, \pi, 4\pi/3, 5\pi/3]$\end{tabular} & $\phi_1,\phi_2\in U[0, 2\pi]$ & $\phi_1,\phi_2\in U [0, 2\pi]$\\
    \hline 
         Mass ratio & $q\in [1,1.5,2,3,4]$ & $q\in U [1,4]$ & $q\in U [1,4]$\\
    \hline 
         Total mass [$M_\odot$] & \begin{tabular}{@{}c@{}}$M\in\left[75,150,250\right]$ \\ (PSD matches only) \end{tabular} & $M\in U [70,250]$ & $M\in [75,100,125,150,175,200,225,250]$ \\
    \hline 
        Inclination & -- & $\cos(\iota)\in U [-1,1]$ & $\iota=\pi/3$\\
    \hline
         Initial phase & -- & $\phi_0\in U [0,2\pi]$ & \begin{tabular}{@{}c@{}c@{}}$\phi_0\in [0,\pi/4, \pi/2, 3\pi/4,$ \\ $\pi, 5\pi/4, 3\pi/2, 7\pi/4]$\\ (included in weighted sum) \end{tabular}\\
    \hline
         Polarization & -- & $\kappa\in U [0,\pi/4
         ]$ & \begin{tabular}{@{}c@{}}$\kappa\in [0,\pi/12, \pi/6,\pi/4
         ]$ \\ (included in weighted sum) \end{tabular}\\
    \hline
         Total binaries sampled & \begin{tabular}{@{}c@{}} White noise matches: 47,136 \\ PSD matches: 141,408 \end{tabular} & 20,833 & ... \\
    \hline
    \hline
    \end{tabular}
    \caption{Binary configurations used in the different match calculations.  Binaries for mode-by-mode match calculations are sampled systematically across the intrinsic parameter space, and three total mass scales are used for the O4 PSD matches. Since it is the relative azimuthal separation of spins which is important, we choose to keep $\phi_1=\pi/3$ fixed while changing $\phi_2$. Additionally, we place constraints on the spin magnitudes and tilt angles such that none of our binary configurations have both BHs with aligned spins or nonspinning. For the strain matches, the intrinsic parameters are drawn from random uniform distributions (shown by $U\left[a,b\right]$ in the table); the SNR-weighted matches use fixed extrinsic parameter values while they are drawn randomly for the sky-and-polarization-averaged matches.}
    \label{tab:MatchParams}
\end{table*}

The mode-by-mode matches are computed for a large number of mass ratios and spins that systematically sample the validity range of the surrogate model with the details provided in the second column of Table~\ref{tab:MatchParams}. We choose the initial time as the reference time, i.e. $\tref\equiv t_0$, and sample the initial spins in a spherical coordinate system using the spin magnitudes $||\vec{\chi}_i||$, the azimuthal orientations $\phi_i=\text{arccos}(\hat{S}_i \cdot \hat{x})$, and the cosine of the tilt angles $\cos(\theta_i) = \hat{S}_i \cdot \hat{L}$. 
Specifically, we keep the initial azimuthal orientation of the spin of the larger BH $\phi_1$ of $\vec{\chi}_{1\perp}$ fixed, while rotating $\vec{\chi}_{2\perp}$ to achieve a range of angular azimuthal separations, and vary the initial tilt angles $\theta_i$ systematically. Further, we only choose configurations with at least one spinning BH, demanding that at least one BH has a nonzero in-plane spin, thereby excluding aligned-spin or nonspinning binaries. This amounts to a total of 47,136 unique binary configurations in terms of their intrinsic parameters $\{q, \vec{\chi}_1, \vec{\chi}_2\}$. When considering a detector PSD, we additionally consider three values of the total mass, $75, 150$ and $250M_\odot$.

For the strain matches as given in Eqs.~\eqref{eq:SkyPolAveragedStrainMatch} and~\eqref{eq:SNRWeightedStrainMatch}, additional extrinsic parameters, namely binary inclination $\iota$, initial phase $\phi_0$, and polarization $\kappa$, need to be taken into account. In such high dimensions, systematic sampling becomes unfeasible. 
Therefore, for the sky-and-polarization-averaged matches we draw the intrinsic and extrinsic binary parameters from random uniform distributions as detailed in the third column of Table~\ref{tab:MatchParams}, considering a total of $20,833$ unique binary configurations. 

For the SNR-weighted strain matches, we first draw 100 binary configurations randomly from the 20,833 used to compute the sky-and-polarization-averaged strain matches, only considering their intrinsic parameters, $\{q,|\vec{\chi}_1|,|\vec{\chi}_2|,\text{cos}(\theta_1),\text{cos}(\theta_2),\phi_1,\phi_2\}$. 
We fix the source inclination at a moderate inclination of $\iota=\pi/3$. As detailed in the last column of Table~\ref{tab:MatchParams}, for each binary configuration we choose eight initial phase and four polarization values and compute 32 matches $\mathcal{M}(h,h_{\sigma})$, one for each pair $\{\phi_0,\kappa\}$, which are then summed into a single SNR-weighted match for each binary configuration as per Eq.~\eqref{eq:SNRWeightedStrainMatch}. We repeat this calculation for each of the total masses detailed in Table~\ref{tab:MatchParams}, noting that we use the same 100 intrinsic binary configurations for each $M$. This yields 800 SNR-weighted strain matches for each mapping $\sigma \in \{\chi_p, \nchip\}$.

\section{Results}
\label{sec:Results}

\subsection{Mode analysis}
\label{sec:modematches}
\begin{figure*}[t!]
\includegraphics[width=\textwidth]{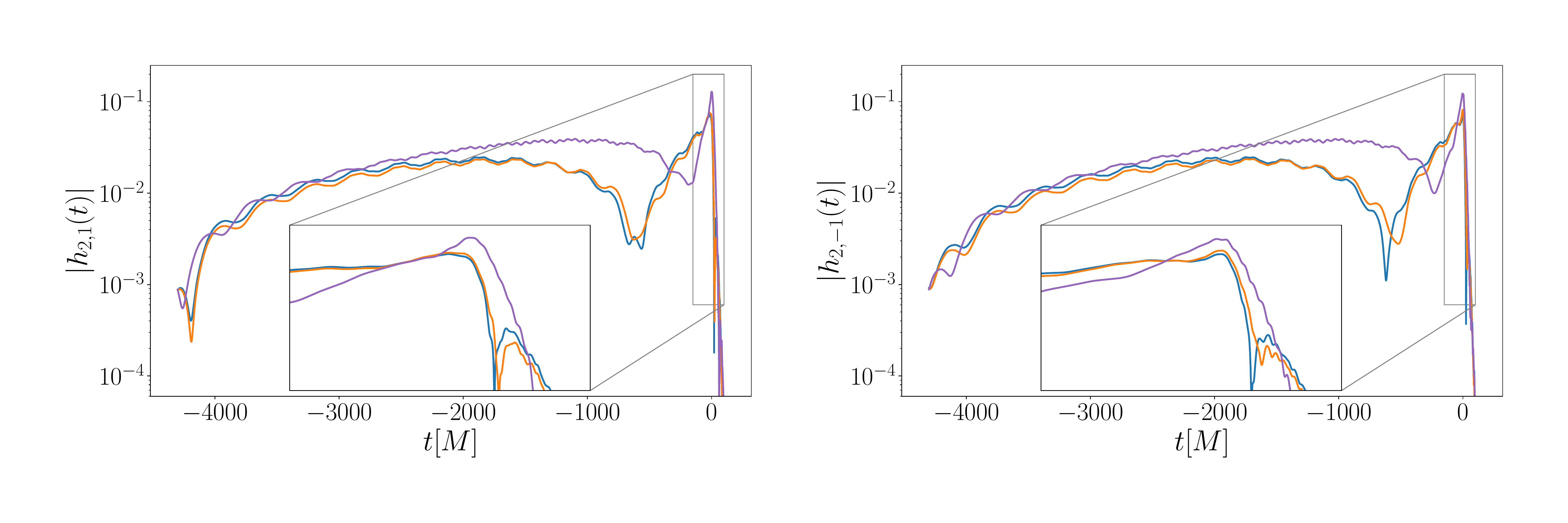} 
\caption{Amplitude of the $(2,1)$-mode (left) and the $(2,-1)$-mode (right) for a fiducial precessing binary with $q=1.4$, $\vec{\chi}_1(t_0)=(0.075,0.043,0.05)$, and $\vec{\chi}_2(t_0)=(-0.346,0.6,-0.4)$. 
The fully precessing signal waveform is shown in blue, and the template waveforms parametrized by $\chi_p$ and $\nchip$ are shown in purple and orange, respectively. The $\nchip$ reproduces the phenomenology of this mode markedly better than the $\chi_p$-mapping, especially in the merger-ringdown portion of the waveform.}
\label{fig:Fiducial}
\end{figure*}

We first assess how well the vectorial effective spin parameter $\nchip$ reproduces individual modes, in particular HMs, for different mass ratios. In Figure~\ref{fig:Fiducial}, we show the $(2,\pm 1)$-modes for a fiducial precessing binary with $q=1.4$ and initial spins $\vec{\chi}_1=(0.075,0.043,0.05)$, $\vec{\chi}_2=(-0.346,0.6,-0.4)$. We find that $\nchip$ (orange) captures the fully precessing modes (blue) significantly better than a simple $\chi_p$ parametrization (purple).
In particular, we see that unlike $\chi_p$, $\nchip$ reproduces the amplitude and phasing of the mode oscillations on the orbital timescale and the amplitude modulations on the precession timescale much more faithfully. Additionally, amplitude modulations in the ringdown signal, which are completely missed in the $\chi_p$-parametrization, are much better captured. We note that the precession of this fiducial binary is dominated by the secondary BH spin -- a region in the spin parameter space where $\chi_p$ knowingly performs poorly. Additional examples are presented in Figs.~\ref{fig:EqualMassFiducial, fig:q3Fiducial, fig:FiducialSwitched} in Appendix~\ref{app:MoreFiducialExamples}, including a mass ratio $q=3$ binary with the precession dominated by the primary BH spin in Figure~\ref{fig:q3Fiducial}, where we still observe noticeable improvements with $\nchip$ over $\chi_p$.

\begin{figure*}[ht!]
\includegraphics[width=\textwidth]{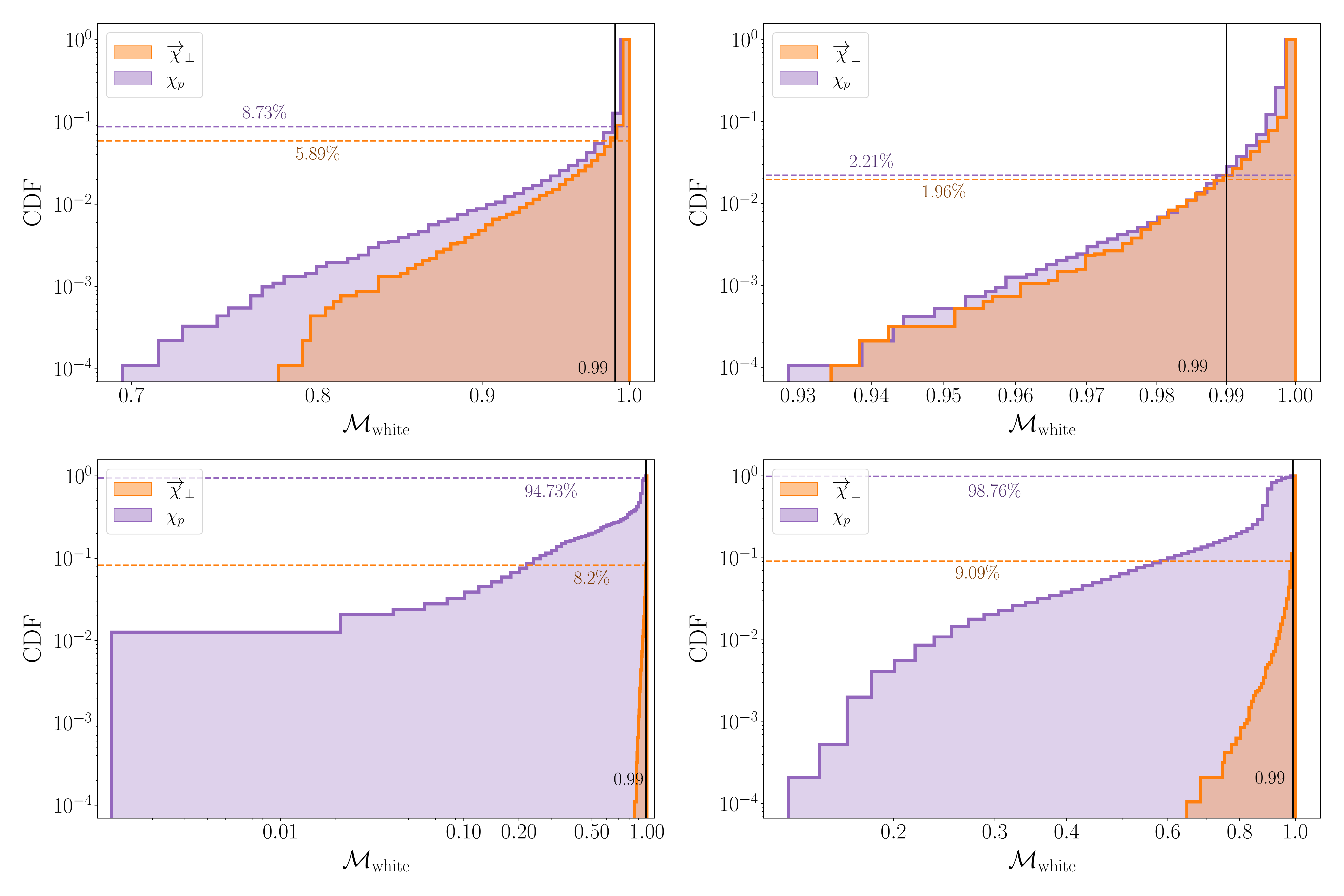}
\caption{Cumulative histograms of white noise matches for the $(2,2)$-mode (top row) and the $(2,1)$-mode (bottom row) for 9,120 binaries with mass ratios $q=1$ (left column) and 9,504 binaries with $q=3$ (right column). Details of how these binaries are systematically sampled can be found in Table~\ref{tab:MatchParams}. Orange histograms show the results using the $\nchip$-parametrization, purple ones $\chi_p$. The dashed horizontal lines indicate the percentage of matches below $0.99$. 
Using $\nchip$, we see a clear improvement over the $\chi_p$-mapping for the $(2,1)$-mode, and comparable if slightly better performance for the $(2,2)$-mode. Results for additional modes and mass ratios are presented in Figure~\ref{fig:FullWhiteMatchResults} in Appendix~\ref{app:FullSamplingandResults}.}
\label{fig:q1whitematchcumul}
\end{figure*}

To quantify the efficacy of $\nchip$ across the parameter space, we compute white noise mode-by-mode matches for the $(2,\pm2)$-modes and a selection of HMs for all binaries listed in the second column of Table~\ref{tab:MatchParams}. Figure~\ref{fig:q1whitematchcumul} shows the cumulative match results for the $(2,2)-$ and $(2,1)$-modes for mass ratio $q=1$ and $q=3$. Results for additional modes and mass ratios are shown in Figure~\ref{fig:FullPSDMatchResults} in Appendix~\ref{app:FullSamplingandResults}.

We expect the $\chi_p$-parametrization to perform well at replicating the dominant $(2,\pm2)$-mode behavior, and indeed we see similar results in this mode for both parametrizations, if slightly improved with the new effective spin, except for the equal-mass case, where we find a more marked improvement. We attribute this to the fact that $\chi_p$ is designed to replicate the average precession rate, but in equal-mass configurations the in-plane spin vectors precess at the same rate and become orientationally locked, which is not captured correctly by $\chi_p$~\cite{Schmidt:2014iyl, Gerosa:2020aiw}. Additionally, $\nchip$ takes into account the in-plane spins on both black holes, while $\chi_p$ selects only the larger spin component leading to a systematic underestimation of the total in-plane spin for equal-mass cases.

We observe the most dramatic improvements in the $(2,\pm1)-$modes. For example, for $q=3$ shown in Figure~\ref{fig:q1whitematchcumul}, the percentage of matches below $0.99$ decreases dramatically from $94.73\%$ with $\chi_p$ to $8.2\%$ with $\nchip$. 
Note that the long tails toward very low matches for the $\chi_p$-parametrization, and the comparatively short ones of $\nchip$, are a generic feature across all HMs we analyzed, suggesting that $\nchip$ better replicates the higher mode behavior even when it performs at its worst.

Intriguingly, for the $(2,\pm1)$ and $(4,\pm4)$-modes, both parametrizations perform worst at $q \sim 1.5$, after which their performance improves with increasing mass ratio. To further investigate this intermediate region between the equal-mass regime and higher mass ratios where the secondary spin becomes less important, we performed additional white noise matches at mass ratios $q\in[1.2,1.4,1.6,1.8]$. We find that the performance of both spin mappings improves with increasing mass ratio for the $(2,\pm2)-$ and $(3,\pm3)$-modes, with $\nchip$ consistently outperforming $\chi_p$. We also find that $\nchip$ performs worst around $q\sim 1.4$ for the $(2,\pm1)-$ and $(4,\pm4)$-modes, but that the distributions for both mappings are fairly flat between $q=1.2$ and $q=2$, and even at its worst $\nchip$ still vastly outperforms $\chi_p$. For example, in the $(2,1)$-mode, at $q=1.4$, the percentage of matches below $0.99$ is $100\%$ with $\chi_p$, and only $26.35\%$ with $\nchip$.
We also note that we find only minor differences between the positive and negative $m$-modes for both mappings, and neither performs consistently worse at replicating either positive or negative $m$-modes. 

Additionally, our new mapping shows moderate improvements for the $(3,\pm3)-$modes and striking improvements for the $(4,\pm4)$-modes, with the improvements particularly marked at equal-mass and at our highest mass ratio $q=4$.
In summary, we find that the $\nchip$-parametrization performs consistently better than $\chi_p$ for every mass ratio and across all modes, and in particular for odd $m$-modes.

In addition to the white noise matches, we repeat the analysis using the projected O4 aLIGO PSD~\cite{NoiseCurves} with $f_{\rm min}=20$Hz for three total masses $M\in[75, 150, 250] M_{\odot}$ compatible with the fixed length of the NR surrogate.
For the PSD mode-by-mode matches, we obtain qualitatively similar results to the white noise matches as shown in Figure~\ref{fig:FullPSDMatchResults} in Appendix~\ref{app:FullSamplingandResults}.
All of the matches improve slightly compared to the white noise matches across both mappings due to the frequency weighting of the PSD, but the features of our results and conclusions remain the same: The $\nchip$-mapping significantly improves upon $\chi_p$ for the $(2,\pm1)$- and $(4,\pm4)$-modes, with moderate improvements for the $(3,\pm3)-$modes, and comparable if slightly better performance for the $(2,\pm2)$-modes.

\subsection{Strain analysis}
\label{sec:precmatches}
\begin{figure}[t!]
\includegraphics[width=\columnwidth]{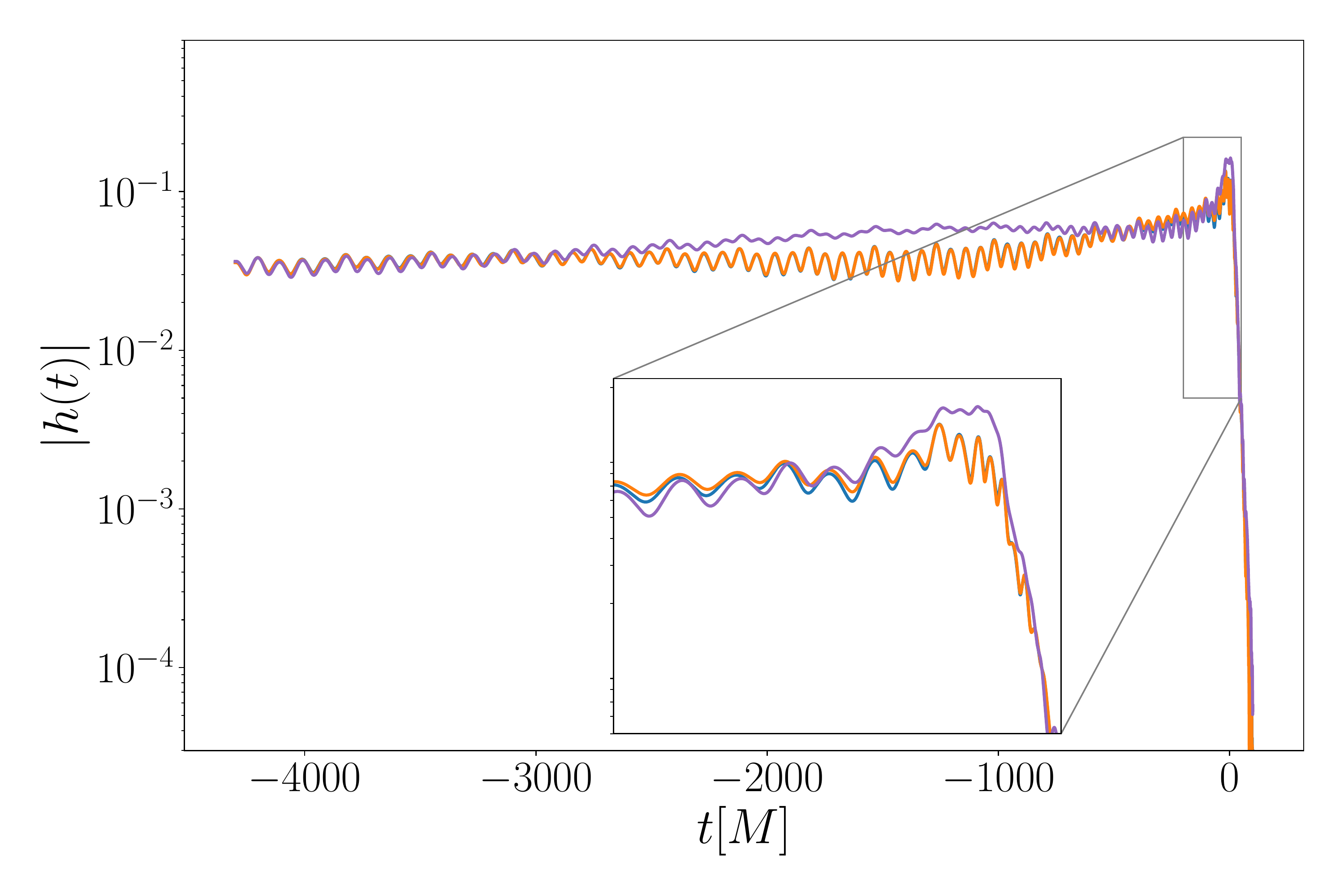}
\caption{Amplitude of the waveform strain $h(t)$ for the same fiducial binary as in Figure~\ref{fig:Fiducial} at an inclination of $\iota=\pi/3$. The figure shows the fully precessing waveform (blue) along with both the $\chi_p$ (purple) and $\nchip$-parametrizations (orange). The $\nchip$-mapping reproduces the strain amplitude much more faithfully than $\chi_p$, especially in the late inspiral portion of the waveform. Note that in the late inspiral, where the blue line cannot be seen, it is indistinguishable from the orange line.}
\label{fig:FiducialStrain}
\end{figure}

\begin{figure}[t!]
\includegraphics[width=\columnwidth]{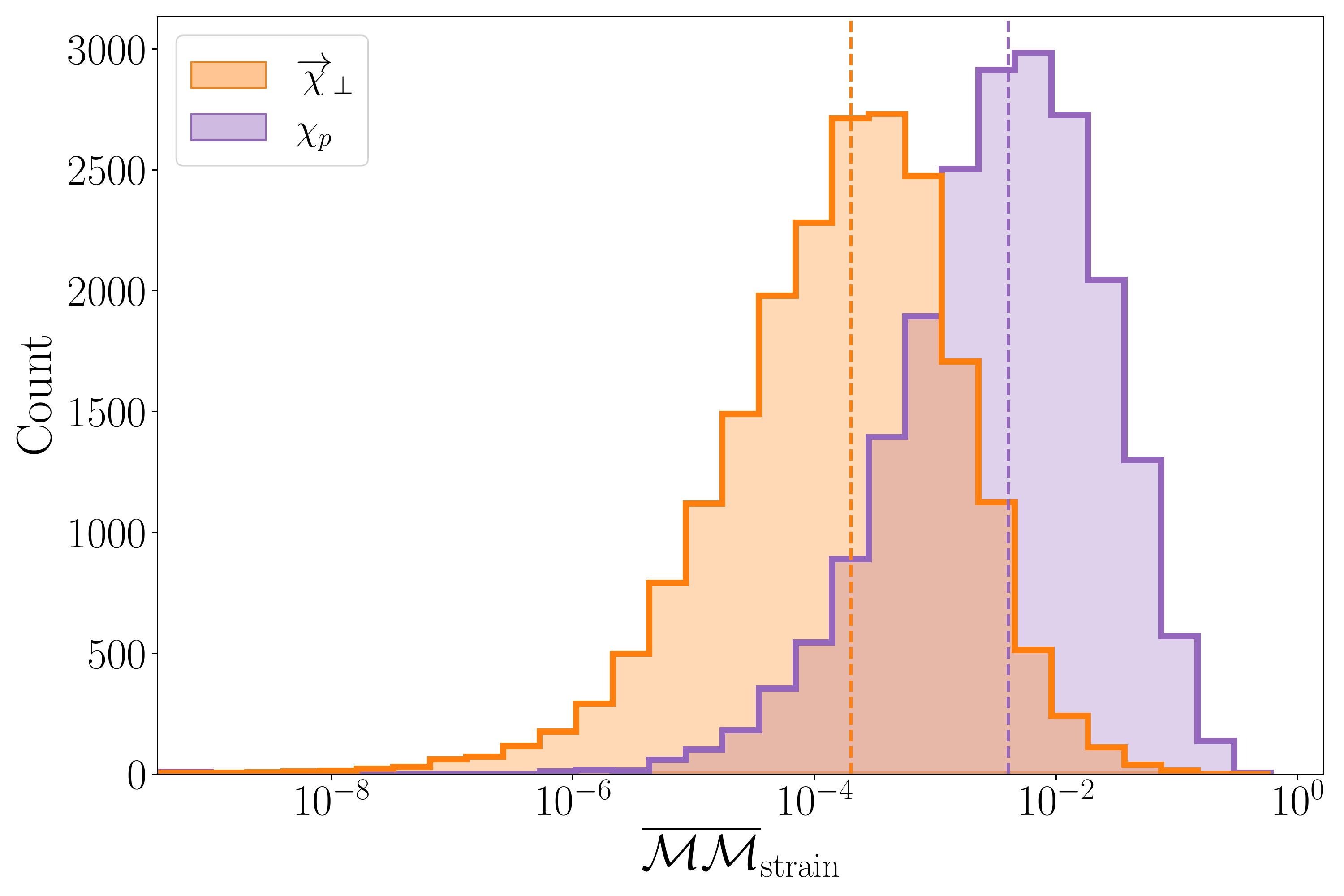}
\caption{Histograms of the sky-and-polarization-averaged strain mismatches $\overline{\mathcal{M}\mathcal{M}}_{\text{strain}}$ between the fully precessing waveform and each of the two-spin mappings using the O4 PSD. The parameters for each of the 20,833 binaries tested were drawn from random uniform distributions as outlined in Table~\ref{tab:MatchParams}. The dashed vertical lines show the median mismatch for each mapping. We see an improvement in the median mismatch of more than one order of magnitude when using $\nchip$.}
\label{fig:strainhist}
\end{figure}

In the previous section we have demonstrated the improvement of $\nchip$ over $\chi_p$ at the level of individual $h_{\ell m}$-modes. We now assess the degree to which the improvement in the HMs impacts the strain. Figure~\ref{fig:FiducialStrain} shows the strain for the fiducial binary at an inclination of $\iota=\pi/3$. The excellent agreement between the fully precessing waveform (blue) and the one parametrized by $\nchip$ (orange) throughout the late inspiral as well as the merger ringdown is clearly visible. To quantify this agreement, we first compute the sky-and-polarization-averaged strain mismatches for 20,833 binary configurations as detailed in Table~\ref{tab:MatchParams} using the O4 PSD and $f_{\rm min}=20$Hz. Our results for both effective parametrizations are shown in Figure~\ref{fig:strainhist}. Using $\nchip$ rather than $\chi_p$, we find a median improvement of more than 1 order of magnitude from $4\times10^{-3}$ to $2\times10^{-4}$. Furthermore, we note the non-negligible tail of extremely low mismatches below $10^{-6}$ for $\nchip$.

\begin{figure*}[t!]
\centering
\includegraphics[width=\textwidth]{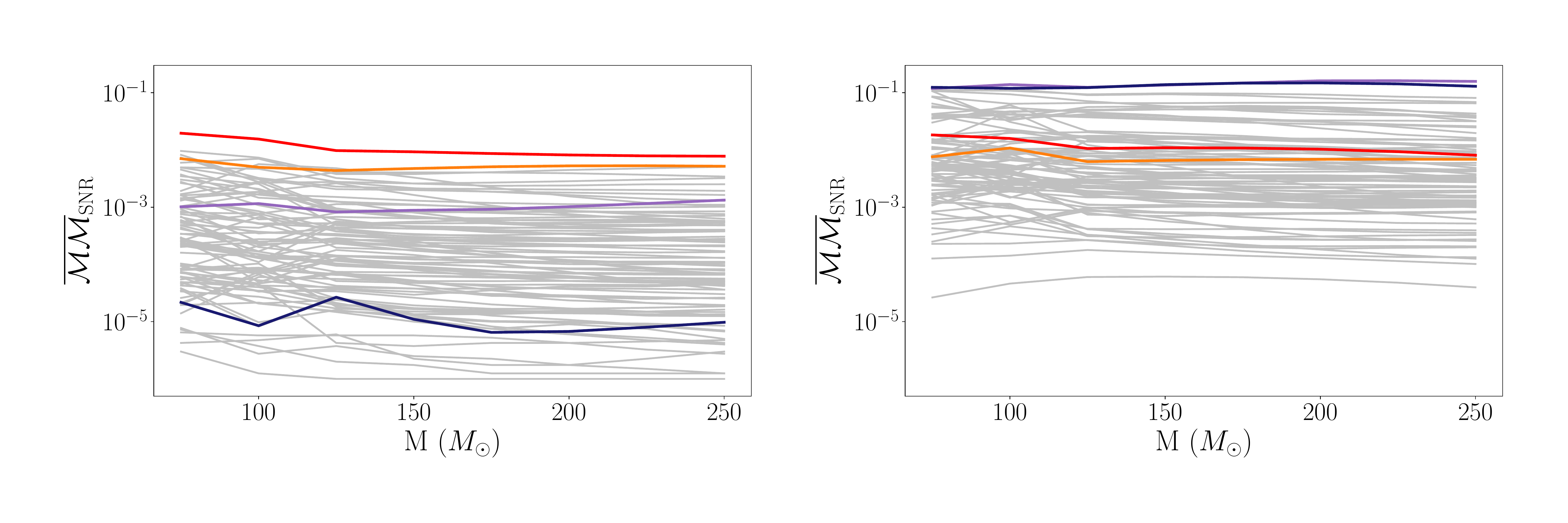}
\caption{SNR-weighted strain mismatches $\overline{\mathcal{M}\mathcal{M}}_{\text{SNR}}$ as a function of binary total mass $M$ for 100 binaries for the $\nchip$ (left) and $\chi_p$ (right) mappings. The red and orange lines correspond to the cases $q=1.3$, $\vec{\chi}_1(t_0)=(0.37,-0.36,0.46)$, $\vec{\chi_2}(t_0)=(-0.32,0.02,0.13)$ and $q=1.6$, $\vec{\chi}_1(t_0)=(-0.21,0.23,0.58)$, $\vec{\chi}_2(t_0)=(-0.26,-0.56,0.34)$ respectively, which show the worst results for the $\nchip$-mapping. The purple and navy lines show the two cases $q=3.2$, $\vec{\chi}_1(t_0)=(-0.66,0.12,0.01)$, $\vec{\chi}_2(t_0)=(0.23,-0.25,-0.03)$ and $q=3.2$, $\vec{\chi}_1(t_0)=(-0.55,0.09,0.07)$, $\vec{\chi}_2(t_0)=(0.03,0.04,0.03)$ where $\chi_p$ shows the worst performance. We see an average improvement of around 1 order of magnitude using the $\nchip$-mapping.}
\label{fig:SNRWeightedStrainMatches}
\end{figure*}

As low matches are often correlated with low SNRs and, therefore, with a lower detection probability, we also compute the SNR-weighted mismatch Eq.~\eqref{eq:SNRMismatch} for 100 randomly drawn intrinsic binary configurations as given in the fourth column of Table~\ref{tab:MatchParams} for a moderate inclination of $\iota=\pi/3$ at $t_0$. Similar to the sky-and-polarization-averaged strain mismatches, we see an improvement of around 1 order of magnitude when using the $\nchip$-parametrization instead of $\chi_p$, as shown in Figure~\ref{fig:SNRWeightedStrainMatches}. 
The worst two cases for each parametrization are highlighted in both panels. We see that the worst cases for $\nchip$ (red and orange) perform similarly under both mappings, if slightly better with the new $\nchip$-mapping. These cases both have a mass ratio of $q\sim1.5$, which as noted previously in Sec.~\ref{sec:modematches}, is a mass ratio where both parametrizations perform worst. The worst cases for $\chi_p$ (purple and blue) on the other hand, show significant improvements of around 2 and 4 orders of magnitude respectively across the entire mass range when the $\nchip$-mapping is used. 

\begin{figure}[ht!]
\includegraphics[width=\columnwidth]{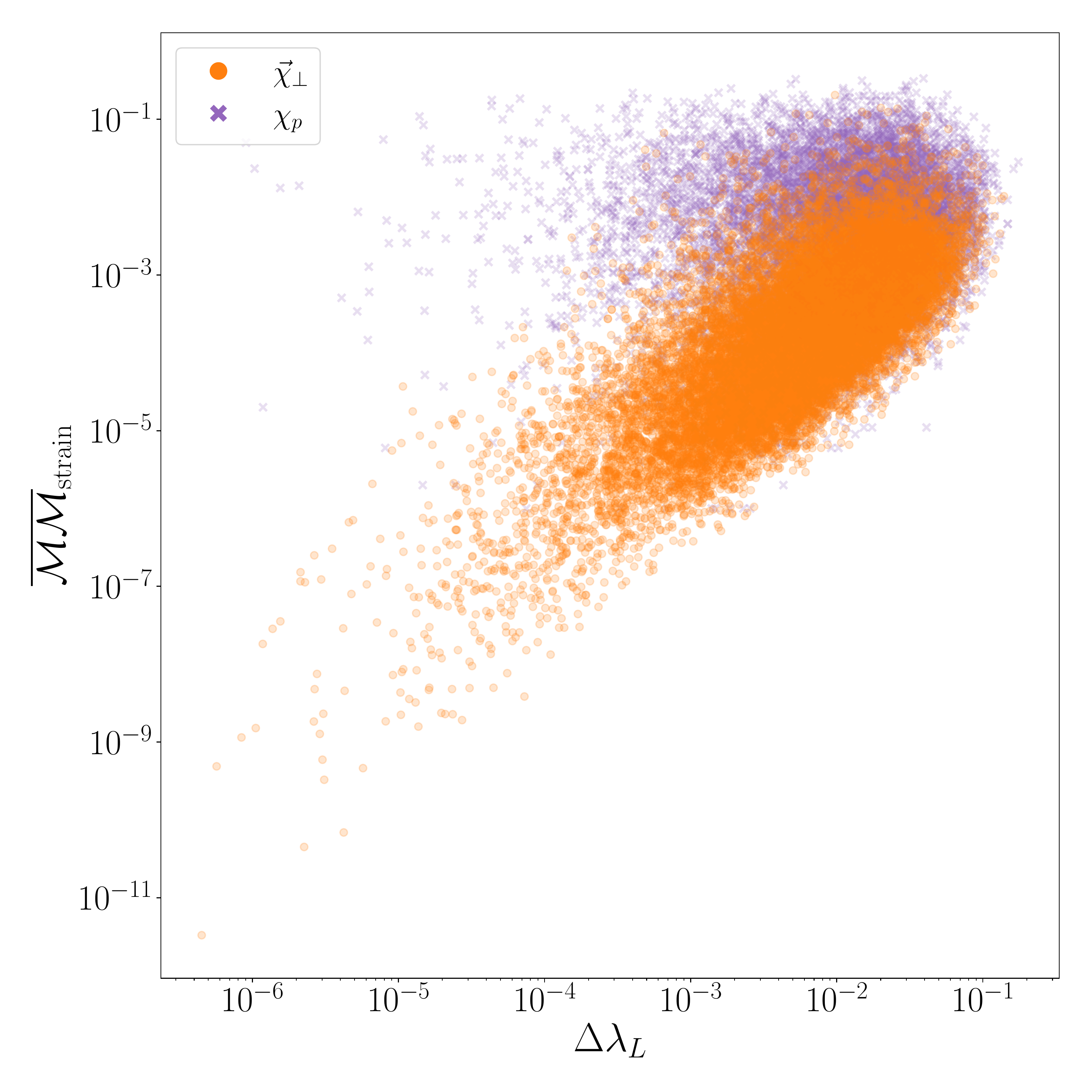}
\caption{Sky-and-polarization-averaged strain mismatch $\overline{\mathcal{M}\mathcal{M}}_{\text{strain}}$ versus $\Delta \lambda_L (t_0)$ for the 20,833 binaries with both the $\nchip$ (orange) and $\chi_p$ (purple) effective spin parametrizations. We observe a slight correlation between small $\Delta \lambda_L (t_0)$ and low mismatches for $\nchip$, yielding a significantly better replication of the initial opening cone in comparison to $\chi_p$.}
\label{fig:dproxy}
\end{figure}

To better understand these marked improvements we employ several diagnostics. First, we investigate whether there exists a correlation between the initial opening angle of the precession cone $\lambda_L(t_0)$ [Eq.~\eqref{eq:cone}] and the sky-and-polarization-averaged strain mismatch. We define the difference in the initial precession cone opening angle between the mapped and unmapped system, $\Delta \lambda_L (t_0)$, as
\begin{align}
\label{eq:proxy}
\Delta \lambda_L (t_0)&\equiv \lambda_L (t_0) - \lambda_{\sigma L} (t_0),
\end{align}
where $\lambda_L$ is given by Eq.~\eqref{eq:cone}. The definition of $\lambda_{\sigma L}$ is the same as for $\lambda_L$, but replaces $S_{\perp}$ with $S_{\sigma \perp}$, where $S_{\perp}$ is the initial total in-plane spin magnitude before the mapping and $S_{\sigma \perp}$ is the total in-plane spin magnitude after the mapping. All quantities are evaluated at the initial time $t_0$, and we approximate $L$ by its Newtonian value $L=\mu \sqrt{Mr}$, where $\mu=m_1 m_2/M$ is the reduced mass and $r=M^{1/3}\omega_{\rm orb}^{-2/3}$ with $\omega_{\rm orb}$ the orbital angular frequency. 

Figure~\ref{fig:dproxy} shows $\Delta \lambda_L (t_0)$ against the strain mismatch for each of the 20,833 binaries, calculated with both the $\nchip$ (orange) and $\chi_p$ (purple) effective spin mappings. 
For the new mapping, we see a clear correlation between lower values of $\Delta \lambda_L (t_0)$ and lower strain mismatches. Overall, the $\nchip$-parametrization yields a more accurate initial cone opening angle resulting in a more faithful representation of the fully precessing waveform.

\begin{figure}[t!]
\includegraphics[width=\columnwidth]{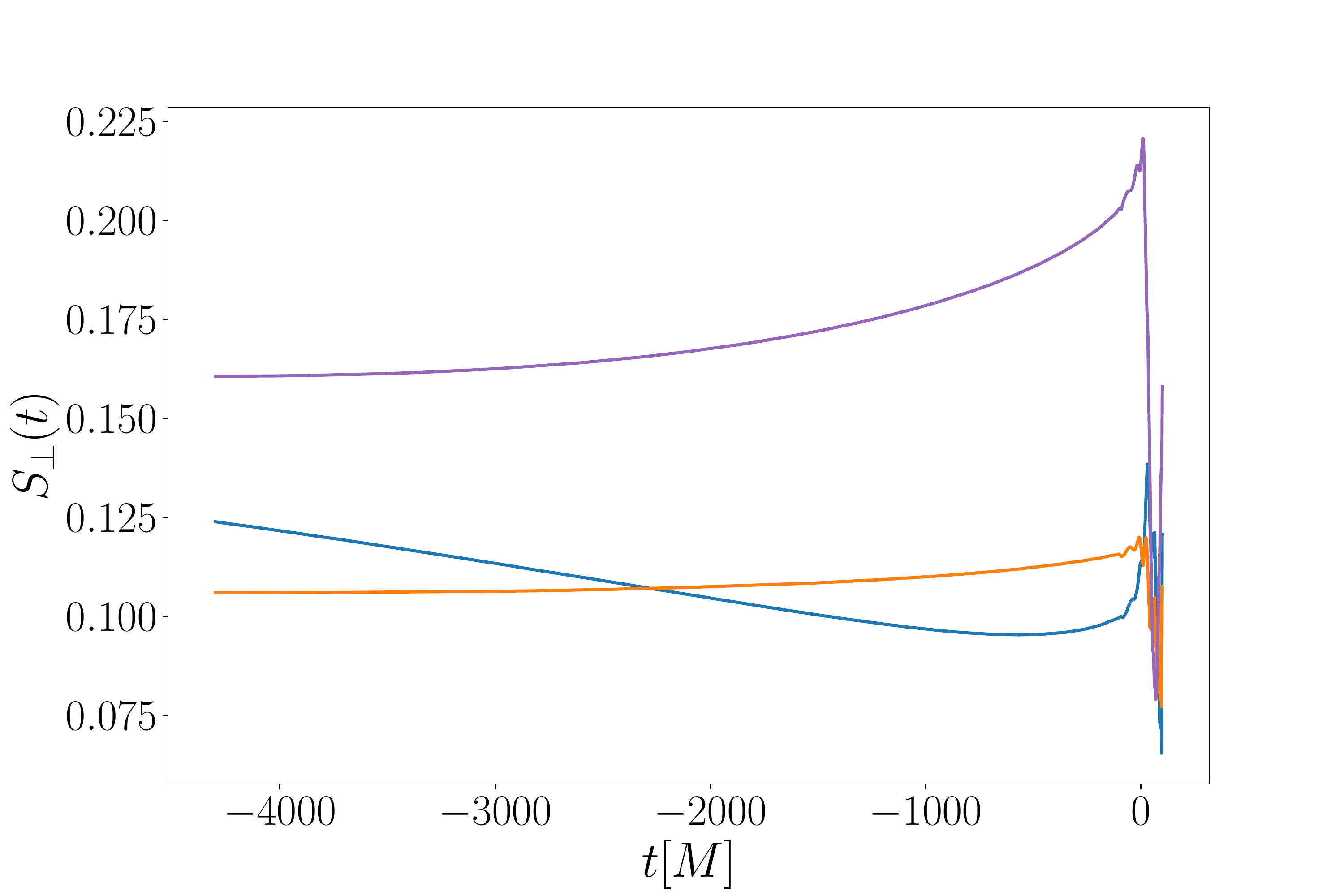}
\caption{Time evolution of total in-plane spin magnitude $S_\perp$ in the coprecessing frame for the same fiducial binary as in Figure~\ref{fig:Fiducial} (blue). The purple graph shows the spin evolution obtained after applying the $\chi_p$-mapping at the initial time $t_0$, the orange graph that of the $\nchip$-parametrization. We see that $\nchip$ preserves the total in-plane spin magnitude, and thus the spin dynamics, much better than the $\chi_p$-mapping.}
\label{fig:FiducialSpinEvolution}
\end{figure}

The better agreement between the initial opening angles suggests that the spins themselves are captured more faithfully. To show this, as a second diagnostic we compare the spin evolutions of a fully precessing binary with its effective counterparts. Figure~\ref{fig:FiducialSpinEvolution} shows the time evolution of the total in-plane spin $S_\perp$ (blue) for the fully precessing fiducial binary and those of the $\nchip$ (orange) and $\chi_p$ (purple) parametrizations for the fiducial binary. We obtain these by transforming the spin evolutions in the inertial frame to the coprecessing frame using the quaternions.
It is evident that the two-dimensional $\nchip$-mapping represents the full-spin dynamics much more faithfully than $\chi_p$.

\begin{figure*}[ht!]
\includegraphics[width=\textwidth]{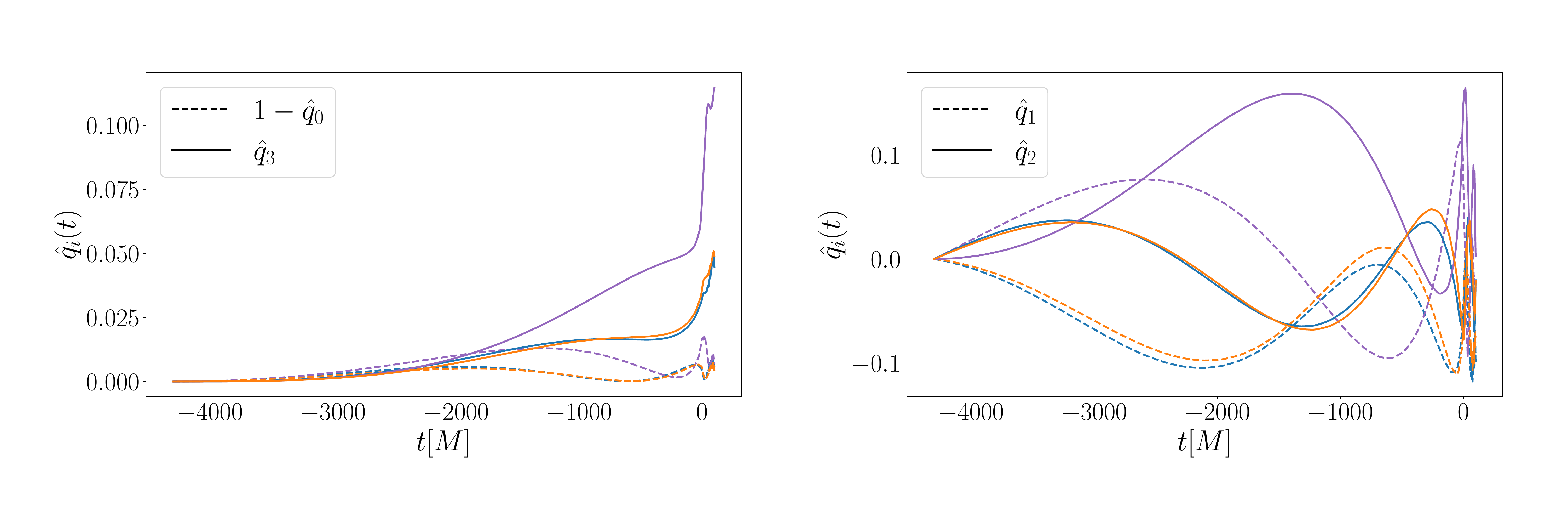}
\caption{Time evolution of the quaternion components $\hat{q}_0$ and $\hat{q}_3$ (left), and $\hat{q}_1$ and $\hat{q}_2$ (right) for the fiducial precessing binary.
The system parametrized by $\nchip$ (orange) reproduces the precession dynamics of the fully spinning system (blue) much more faithfully than the $\chi_p$-mapping (purple).}
\label{fig:quaternions}
\end{figure*}

As a third diagnostic, we investigate how faithfully both mappings reproduce the fully spinning precession dynamics, which is represented by the unit quaternions $\hat{q}_i$. In Figure~\ref{fig:quaternions} we show the time evolution of the four unit quaternion components of the fiducial binary. The new effective spin mapping $\nchip$ clearly replicates the time evolution of each quaternion component much more accurately than $\chi_p$, with the most dramatic improvement observed for $\hat{q}_1$ and $\hat{q}_2$. 

To quantify the improvement in replicating the precession dynamics, we perform a match calculation for each of the four quaternion components $\hat{q}_i$, $i\in[0,1,2,3]$, similar to the white match calculation in Eq.~\eqref{eq:Match} but replacing the waveforms $h$ and $h_{\sigma}$ with the quaternion components, 
\begin{equation}
    \mathcal{M}_{\hat{q}_i} = \mathcal{M}(\hat{q}_i,\hat{q}_{\sigma,i}),
\end{equation}
where we use $S_n = 1$, $\hat{q_i}$ denotes the quaternion component from the fully precessing system, and $\hat{q}_{\sigma,i}$ is the quaternion component produced by the effective mapped system, with $\sigma\in[\chi_p, \nchip]$.

\begin{figure*}[ht!]
\includegraphics[width=\textwidth]{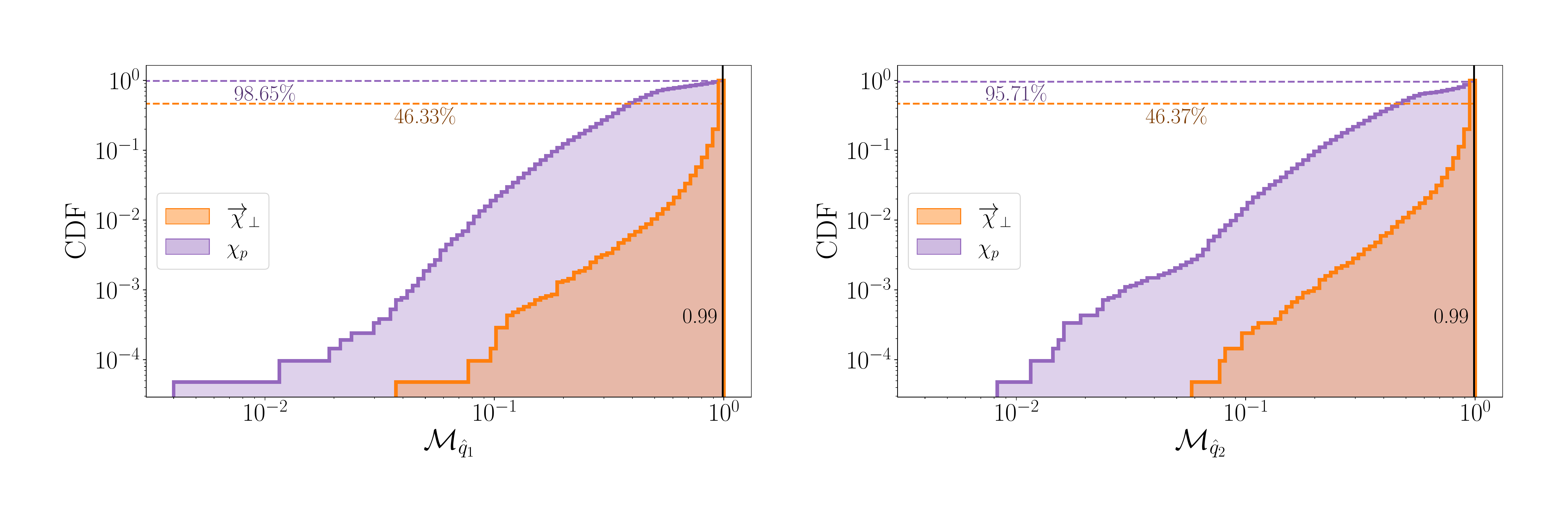}
\caption{Cumulative distribution of matches for two of the four quaternion elements $\hat{q}_1$ (left) and $\hat{q}_2$ (right), between the fully precessing dynamics and each of the $\nchip$ (orange) and $\chi_p$-mapped (purple) systems. The dashed horizontal lines indicate the percentages of matches which are below $0.99$ for each effective mapping. We see significant improvements for the $\nchip$-parametrization over $\chi_p$.}
\label{fig:quaternionshist}
\end{figure*}

We compute the quaternion matches for the same 20,833 binaries used in the sky-and-polarization-averaged strain match calculations. Figure~\ref{fig:quaternionshist} shows the results for $\hat{q}_1$ (left) and $\hat{q}_2$ (right), which show the largest improvements: The percentage of matches below $0.99$ improves from $98.65\%$ with $\chi_p$ to $46.33\%$ with $\nchip$ for $\hat{q}_1$, and for $\hat{q}_2$ it improves from $95.71\%$ to $46.37\%$. 
We see a negligible difference in the results for $\hat{q}_0$, which is well reproduced by both spin mappings: None of cases have a match value below $0.99$. We see a small improvement in the results for $\hat{q}_3$, with the percentage of quaternion matches below $0.99$ dropping from $40.5\%$ for $\chi_p$ to $34.21\%$ for $\nchip$. These results indicate that the observed improvements when using $\nchip$ can indeed be attributed to a more faithful representation of precession dynamics itself.

\subsection{Accuracy of the final spin and recoil}
\label{sec:final}
Finally, we quantify how well the $\nchip$-parametrization is able to reproduce the final spin and recoil of the remnant black hole. The final mass and spin of the remnant determine the quasinormal modes of the ringdown \cite{Press:1971wr,Chandrasekhar:1975zza,Detweiler:1980gk,Kokkotas:1999bd}, so it is therefore crucial to understand the accuracy with which the final state can be replicated by the reduced set of spin parameters. 
We will focus on the final spin estimates as previous comparisons against NR simulations have shown that the final mass estimate is only very weakly dependent on precession~\cite{LALPv2}.

To evaluate the final spin using the surrogate model we first evolve the BH spins in the inertial frame from $t_0$ to a time $t=-100M$ before merger, which are then used to evaluate the remnant fits of Ref.~\cite{Varma:2018aht} via the public \textsc{Python} package \texttt{surfinBH}~\cite{surfinBH}. 
The same procedure is followed to obtain the results under the two-spin parametrizations, where either effective spin map is applied at the initial time $t_0$.
We evaluate the remnant spin for the 20,833 binary configuration of column two in Table~\ref{tab:MatchParams}. We assess the accuracy of the final state under the two mappings by calculating the differences in the remnant spin magnitude $\Delta \chi_f$, the final spin tilt angle $\Delta \theta_f$, the azimuthal spin angle $\Delta \phi_f$, the recoil velocity $\Delta v_f$ and its tilt angle $\Delta \theta_{v_f}$ defined as
\begin{align}
    \Delta \chi_f &= ||\vec{\chi}_f|| - ||\vec{\chi}_{f\sigma}||, \\
    \Delta \theta_f &= ||\arccos(\hat{z}\cdot\hat{\chi}_f)-\arccos(\hat{z}\cdot\hat{\chi}_{f\sigma})||, \\
    \Delta \phi_f &= \arccos(\hat{\chi}_{f\perp} \cdot \hat{\chi}_{f\sigma\perp}), \\
    \Delta v_f &= ||\vec{v}_f|| - ||\vec{v}_{f\sigma}||, \\
    \Delta \theta_{v_f} &= ||\arccos(\hat{z}\cdot\hat{v}_f)-\arccos(\hat{z}\cdot\hat{v}_{f\sigma})||,
\end{align}
where $\sigma\in[\chi_p,\nchip]$, and $\vec{\chi}_{f\perp}$ indicates the $xy$-components of the final spin vector in the inertial frame. 
We note that the remnant spin and recoil velocities are also returned in the inertial coordinate frame of the NR surrogate, which has no particular physical meaning postmerger. However, as we are computing relative differences in magnitudes and angles, this gauge choice has no effect on the results presented here.

\begin{figure}[ht!]
\includegraphics[width=\columnwidth]{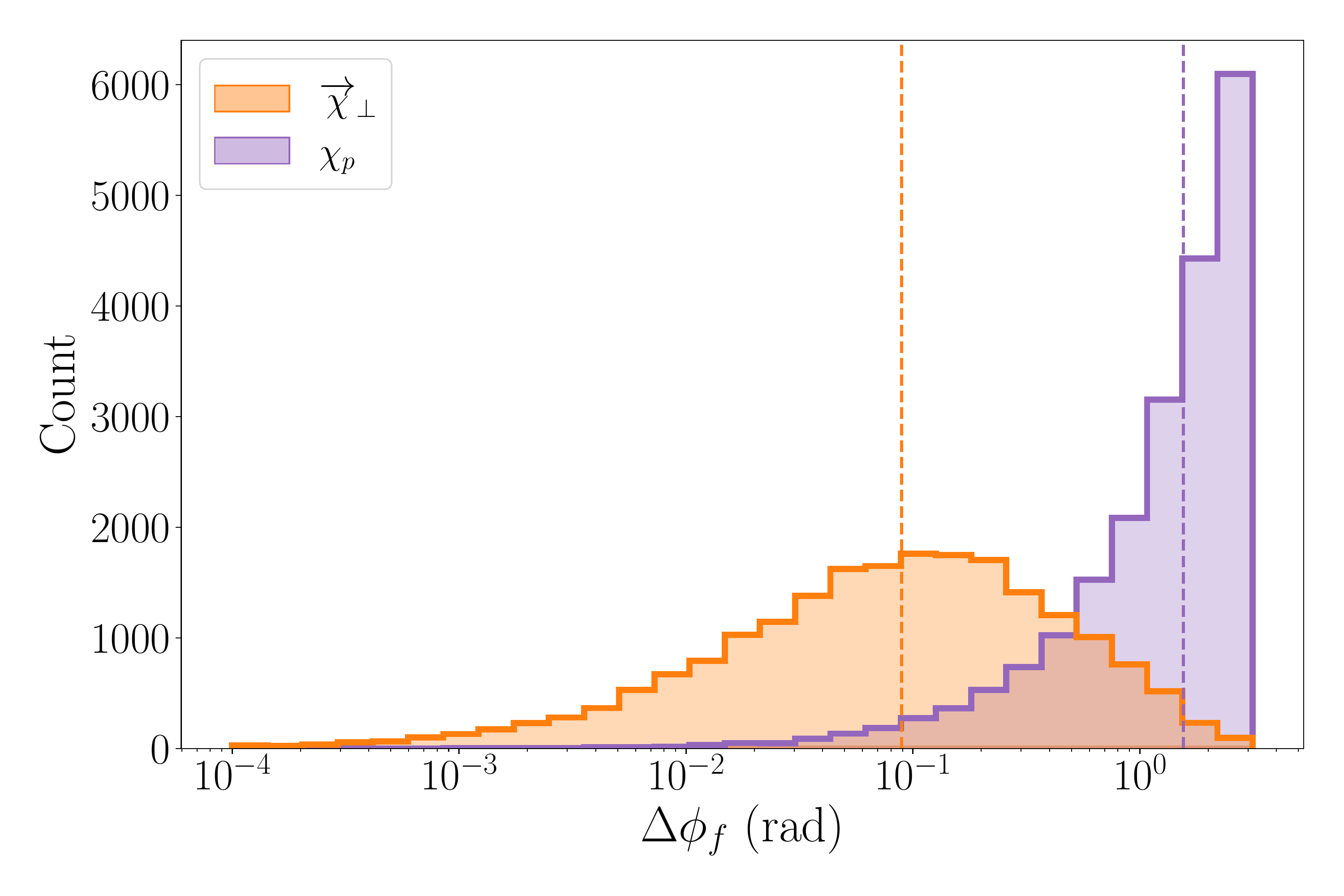}
\caption{Error in the azimuthal angle of the final spin state, $\Delta \phi_f$, in radians, between the final spin state produced by the fully precessing waveform, $\chi_f$, and the resulting final spin state of the waveform produced by the $\nchip$ (orange) and $\chi_p$ (purple) mappings. The new effective spin $\nchip$ reproduces the azimuthal angle of the remnant spin much more accurately, reducing the median error to less than 0.1 rad.}
\label{fig:qFinalStateAng}
\end{figure}

We find marginal improvements in the accuracy of the final spin magnitude and tilt angle using the $\nchip$-mapping as opposed to $\chi_p$. 
The median tilt angle difference $\Delta \theta_f$ improves slightly from $2.30\times 10^{-3}$ rad with $\chi_p$ to $1.43\times 10^{-3}$ rad with $\nchip$; the absolute value $\Delta \chi_f$ also improves slightly from $1.75\times 10^{-3}$ for $\chi_p$ to $9.96\times 10^{-4}$ for $\nchip$.
However, the largest improvement is found for the azimuthal angle $\Delta \phi_f$, which encapsulates the difference in the relative angle in the $xy$-plane of the inertial frame as shown in Figure~\ref{fig:qFinalStateAng}. We see a dramatic difference between the two mappings, with $\nchip$ effectively reproducing the azimuthal orientation with a median error of less than $0.1$ rad, whereas the $\chi_p$-mapping poorly replicates the azimuthal orientation with a median difference of more than 1 rad, and a significant proportion of differences around $\Delta \phi = \pi$. We also note the significantly long tail of the $\nchip$ histogram toward angle differences of zero. 

\begin{figure*}[ht!]
\includegraphics[width=\textwidth]{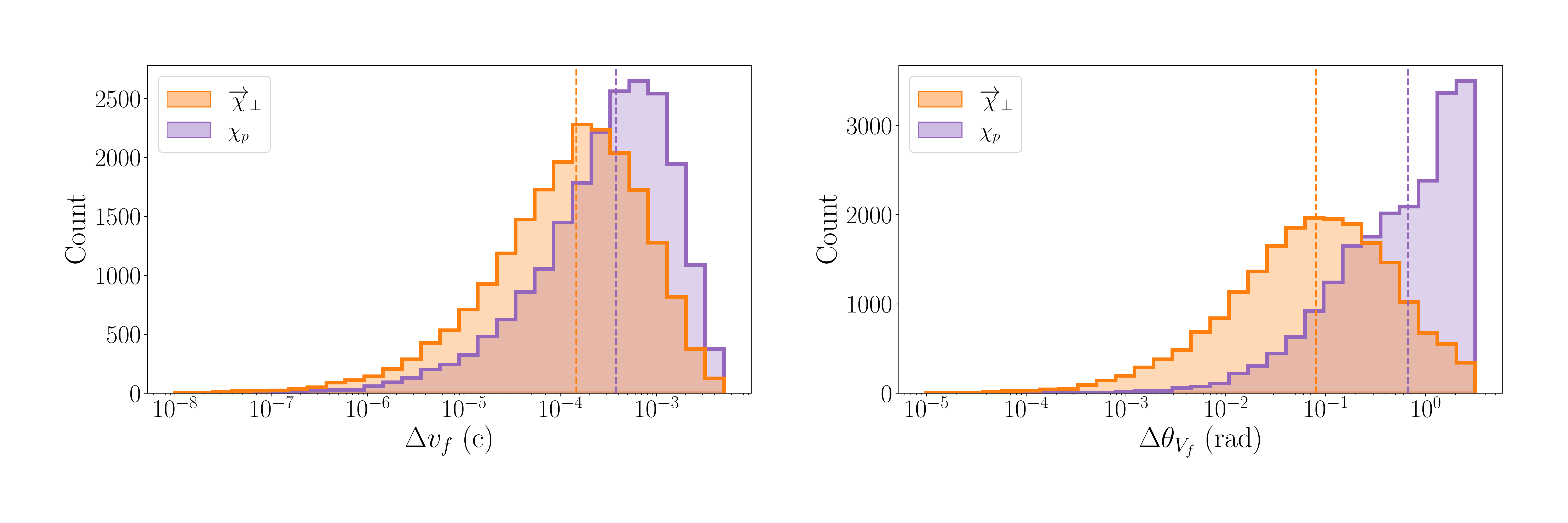}
\caption{Absolute value of the error in the recoil velocity magnitude $\Delta v_f$ (left) in units of c, and recoil velocity tilt angle $\Delta \theta_{v_f}$ (right) in radians, between the fully spinning waveform and each of the $\nchip-$ (orange) $\chi_p-$ (purple) mappings. The dashed vertical lines indicate the median error values.}
\label{fig:RecoilVeloc}
\end{figure*}

We now analyze the effect of the two mappings on the recoil velocity $v_f$ of the final black hole. For the recoil velocity tilt angle $\Delta \theta_{v_f}$, i.e. the polar direction of the recoil, we find a large improvement from a median error of $0.67$ rad for $\chi_p$ to $0.08$ rad for $\nchip$ as shown in the right panel of Figure~\ref{fig:RecoilVeloc}. For the recoil velocity itself, we only find a modest improvement in $\Delta v_f$ from a median error of $3.81\times10^{-4}\, c$ for $\chi_p$ to $1.47\times10^{-4}\, c$ for $\nchip$, corresponding to an improvement in accuracy of $\sim 70$ km/s on average.

To summarize, overall the $\nchip$-parametrizations reproduce the final state, in particular the orientation of the final spin and the direction of the recoil, much more accurately. Both effective spin parametrizations perform similarly in determining the final spin magnitude.

We note that the comparison using $\chi_p$ is not directly comparable to the definition of the final spin used in semianalytical waveform models, which use (a variety of) in-plane spin corrections to modify the final spin of an aligned-spin binary~\cite{Pratten:2020ceb}.
The current generation of precessing Phenom models~\cite{Hannam:2013oca, Khan:2019kot, Pratten:2020igi} apply a correction of the form $S_p/M_f^2$, where $M_f$ is the remnant mass and $S_p$ is an effective in-plane spin contribution. In \cite{Hannam:2013oca,Khan:2019kot}, $S_p$ is taken to be defined as in Eq.~\eqref{eq:s_p def}, which is similar to the results presented here obtained by applying the $\chi_p$-mapping~\cite{LALPv2, Jimenez-Forteza:2016oae}. For the more recent model presented in~\cite{Pratten:2020ceb}, a range of different final spin mappings have been implemented including the $\chi_p$-mapping as well as a precession-averaged mapping that attempts to account for the change in the aligned-spin components due to nutation effects. The EOB models \cite{Ossokine:2020kjp} employ the final spin fits of \cite{Hofmann:2016yih} which introduce corrections to the aligned-spin final state fits that depend on the angle between the two in-plane spin vectors and the projection of the spins along the orbital angular momentum. As discussed in \cite{Ossokine:2020kjp, Chandrasekhar:1975zza} and above, a crucial choice is the separation at which the spins are used to evaluate the final state, taken to be $r = 10M$ in \cite{Ossokine:2020kjp}. This approach enables the effective-one-body models to account for the evolution of the spin vectors ensuring that the same waveform is produced irrespective of the initial separation. 

\section{Discussion}
\label{sec:Discussion}
The inclusion of fully relativistic precession effects in semianalytic IMR waveform models in the strong-field regime remains a challenging problem, with none of the current waveform models from either the Phenom or the EOB waveform family including calibration to NR in the precessing sector. The high dimensionality of the precessing BBH parameter space obfuscates a clear path for calibration. Effective spin parametrizations to reduce the number of spin degrees of freedom are a promising way forward to including fully relativistic precession in the strong-field regime. 
Previously, a scalar quantity $\chi_p$ was introduced to this effect but its efficacy was only demonstrated for the inspiral regime~\cite{Schmidt:2014iyl}. Here, we have assessed its applicability in the strong-field regime.
Crucially, we have found that while $\chi_p$ does faithfully represent the $(2,2)$-mode of the majority of fully precessing systems, it does not accurately reproduce HMs. Since HMs are excited by mass and spin asymmetries, which can be very pronounced for precessing binaries, they carry crucial parameter degeneracy breaking power~\cite{Varma:2014jxa,Graff:2015bba,Bustillo:2015qty,Bustillo:2016gid,Harry:2017weg,Colleoni:2020tgc} making the accurate modeling of HMs critical. Therefore, NR calibration through a simple $\chi_p$-parametrization is unlikely to be sufficient to satisfy the accuracy requirements for future GW observations.

To improve upon the shortcomings of $\chi_p$, we have introduced a new two-dimensional effective precession spin vector, $\nchip$, and have performed extensive studies comparing the efficacy of $\nchip$ to that of $\chi_p$ in the strong-field regime using the NR surrogate waveform model \texttt{NRSur7dq4}~\cite{Varma:2019csw}. 
When analyzing individual $h_{\ell m}$-modes, in particular the $(2,\pm1)$-modes, we have found that $\nchip$ performs significantly better than $\chi_p$, but both effective parametrizations yield comparable results for the quadrupolar $(2,\pm 2)$-modes. Correspondingly, we also have found a significant improvement in the precessing strain matches with the new mapping, from which we have concluded that the improved efficacy of $\nchip$ over $\chi_p$ for the HMs has a significant effect on the accuracy of the overall strain, demonstrating the importance of accurately modeling HMs in precessing systems. 
Furthermore, we have found that $\nchip$ performs better compared to $\chi_p$ in the equal-mass limit (see Figure~\ref{fig:q1whitematchcumul}). In this limit, the BH spins precess at the same rate, locked in orientation relative to each other. The parameter $\chi_p$, which is defined to mimic the average rate of precession, performs knowingly poorly in this limit~\cite{Schmidt:2014iyl}; $\nchip$, on the other hand, is constructed such that it approximates the total in-plane spin of the fully precessing system at some reference time, leading to a significantly improved behavior in the equal-mass limit as anticipated from PN theory~\cite{Apostolatos:1994mx,Buonanno:2004yd}. 
As expected, we have found that $\nchip$ performs increasingly better for larger mass ratios $q\gtrsim2$, where the spin on the smaller BH becomes negligible and hence the approximation with a single in-plane spin becomes more accurate. However, in the intermediate region between these two regimes, whilst still a considerable improvement upon $\chi_p$, we have found a small drop in accuracy at a mass ratio of $q\sim 1.4$, where two-spin effects are important but are not fully captured in $\nchip$.

We have further demonstrated that the overall improvement relative to a $\chi_p$-parametrization can be attributed to a more accurate replication of the precession dynamics itself when using the $\nchip$-parametrization. Indeed, in the case where only one of the two objects has nonzero in-plane spin components, the full dynamics are returned exactly, which is not the case for $\chi_p$. The accurate capture of the precession dynamics of particular interest as a natural way for incorporating strong-field precession information into waveform models is through calibrating the precession dynamics itself, i.e. the rotation operator $\mathbf{R}$ of Eq.~\eqref{eq:hP} or, equivalently, the quaternions. 

Additionally, we have also quantified how well the $\nchip$-mapping is able to replicate the final spin and recoil velocity of the remnant black hole. We have found a considerable improvement in the accuracy with which we have replicated the azimuthal direction of the remnant spin, and moderate improvements in the accuracy of the magnitude and direction of the recoil velocity. 
This suggests our $\nchip$-mapping is better able to replicate the final direction of GW emission, compared to $\chi_p$. Previous work has demonstrated that the relative orientation of the in-plane spins at merger plays a crucial role in determining the final state properties~\cite{Gonzalez:2006md, Brugmann:2007zj, Campanelli:2007cga, Baker:2008md}. We have attributed the observed improvements to the (partial) incorporation of two-spin effects, which are crucial for determining the recoil direction and velocity of the final BH. 

Despite its significantly better performance in all areas, there are also caveats associated with $\nchip$: 
(i) For spin configurations with similar in-plane spin magnitudes, i.e. $S_{1\perp} \simeq S_{2\perp}$, we expect larger mismatches due to, by construction, the neglect of larger in-plane spin-spin couplings. 
(ii) We have normalized $\nchip$ such that the Kerr limit is not violated. Consequently, for binaries with large spin magnitudes, $\nchip$ will underestimate the magnitude of the in-plane spin in the system. Due to the limited spin parameter range of the surrogate, we have not been able to fully quantify the effect of this on the performance of the mapping. 
(iii) The conditional placement of $\nchip$ on either of the two black holes introduces a discontinuity, in the sense that waveforms with $\nchip$ placed on the primary BH show slightly different features from those with $\nchip$ on the secondary BH. We note that all of our $\nchip$-mapped individual waveforms are physical and continuous, but that a shift in phenomenological features can occur between binary configurations with $S_{1\perp}=S_{2\perp}+\epsilon$ where $\nchip$ is placed on the primary BH, against the same binary configuration with slightly smaller $S_{1\perp}=S_{2\perp}-\epsilon$, where $\nchip$ will be placed on the secondary BH. With this in mind, we have tested the performance of $\nchip$ without the conditional placement. We have recalculated the sky-and-polarization-averaged strain matches shown in Figure~\ref{fig:strainhist} with $\nchip$ always placed upon the primary BH irrespective of whether the precession is dominated by the primary or secondary BH, and indeed have found little difference from the original $\nchip$ strain match distribution, with the median mismatch increasing minimally from $2\times10^{-4}$ to $2.08\times10^{-4}$. Additionally, we have recalculated the white noise mode-by-mode matches of Eq.~\eqref{eq:Match}, again with $\nchip$ always placed on the primary BH. While we have found little difference between the $(2,\pm 2)$-mode results for $\nchip$ with and without conditional placement, we have found that it has a marked effect on the results for HMs. For example, in the $(2,1)$-mode at mass ratio $q=3$, the percentage of mismatches below $0.99$ using $\nchip$ without conditional placement rises to $41.9\%$, compared to just $9.1\%$ if we include the conditional placement (under the $\chi_p$-mapping the value is $98.8\%$). We therefore have concluded that for HMs, it is crucial to accurately capture spin asymmetries by placing the effective spin appropriately, to achieve an accurate mapped waveform mode. 

Lastly, we have also tested whether the improvements found by using $\nchip$ over $\chi_p$ are entirely due to the conditional placement, and whether an analogous conditional placement of $\chi_p$ would have similar effects. An example of imposing this condition also on $\chi_p$ is shown in Figure~\ref{fig:FiducialSwitched} in Appendix~\ref{app:MoreFiducialExamples} for the fiducial binary, and we have indeed seen that the phenomenology is captured better. To quantify the improvement in the performance of $\chi_p$ when imposing conditional placement, we have recalculated the sky-and-polarization-averaged strain matches shown in Figure~\ref{fig:strainhist} with an analogous conditional placement for $\chi_p$. We have found only a small improvement compared to the $\chi_p$-mapping without conditional placement, with the median strain mismatch improving from $4\times 10^{-3}$ to $3.4\times10^{-3}$, compared to a median of $2\times10^{-4}$ with $\nchip$. We have also recalculated the mode-by-mode white noise matches shown in Figure~\ref{fig:q1whitematchcumul}, for both effective spin parametrizations, with and without conditional placement. These results are shown in Figure~\ref{fig:ExtraWhiteMatches} in Appendix~\ref{app:MoreFiducialExamples}. We see that for HMs at unequal-mass ratios, neither a conditionally placed $\chi_p$, nor $\nchip$ always placed on the primary object, can replicate the dramatic improvements we have previously seen in Figure~\ref{fig:q1whitematchcumul}. We therefore have surmised that the improvements we have seen in the performance of $\nchip$ over $\chi_p$ are due to a combination of both the new parametrization itself, and the conditional placement, and that both are a necessary requirement to reproduce accurate precessing higher-order waveform modes. 

Finally, we also note that the efficacy of $\nchip$ has not been investigated for the special case of transitional precession, which leads to the tumbling of the total angular momentum $\hat{J}$ when $L \simeq S$ and $\hat{L} = -\hat{S}$. As with all effective mapping that neglects some spin contributions, however, we expect that the fine-tuned conditions needed for the occurrence of the transitional precession phase are not preserved under the mapping.  

In conclusion, our results have demonstrated that by introducing the two-dimensional vector quantity $\nchip$, which partially accounts for two-spin effects, we can accurately reproduce the waveforms of fully precessing binaries, and in particular their HMs, in the strong-field regime across a wide range of the BBH parameter space. The effective reduction of four in-plane spin components to two provides a clear and tractable path forward to meaningfully incorporating precession effects in the strong-field regime into semianalytic waveform models with HMs, which we leave to future work.

\section*{Acknowledgments}
We thank Serguei Ossokine, Sascha Husa, Davide Gerosa, Juan Calderón Bustillo, Matthew Mould, Daria Gangardt and our anonymous reviewer for useful discussions and comments. L. M. T. is supported by STFC, the School of Physics and Astronomy at the University of Birmingham and the Birmingham Institute for Gravitational Wave Astronomy. P. S. acknowledges the Dutch Research Council (NWO) Veni Grant No. 680-47-460. This manuscript has the LIGO document number P2000509.

\appendix

\begin{figure*}[th!]
\centering
\includegraphics[width=0.8\textwidth]{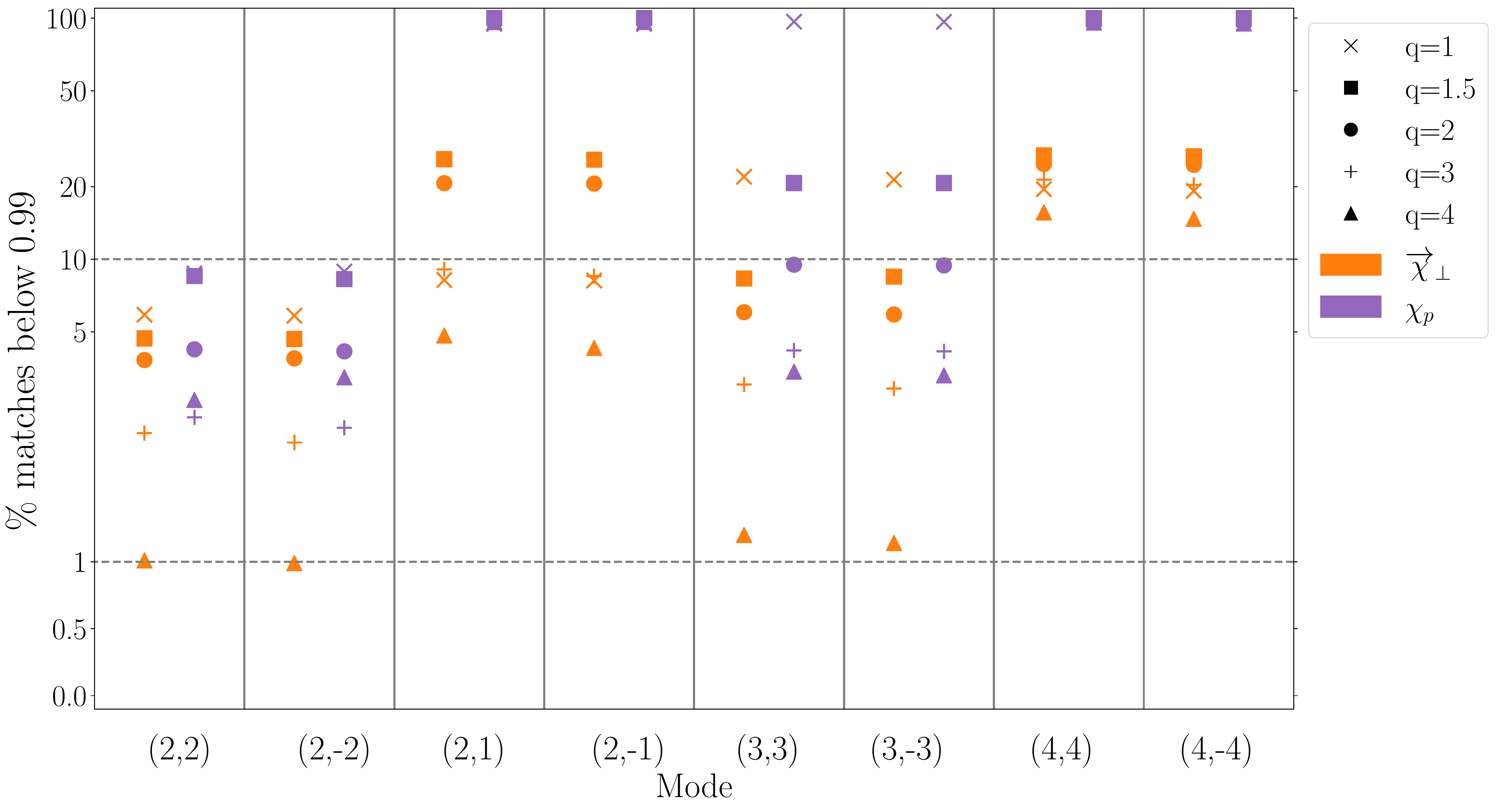}
\caption{Complete results for white noise matches between the fully spinning waveform and each of the effective spin parametrizations, $\nchip$ (orange) and $\chi_p$ (purple). We show percentages of matches split by mass ratio and mode, which fall below a threshold of $0.99$.The dashed horizontal lines indicate the $1\%$ and $10\%$ marks. We see improved results using $\nchip$ as opposed to $\chi_p$ across all mass ratios and modes, and note particularly the dramatic improvements in performance for HMs, especially the $(2,\pm1)$-modes.}
\label{fig:FullWhiteMatchResults}
\end{figure*}

\section{Complete Mode-by-Mode Match Results}
\label{app:FullSamplingandResults}
Here we present the complete results of the white noise and O4 aLIGO PSD matches, for which a selection was presented in Figure~\ref{fig:q1whitematchcumul}. As described in Sec.~\ref{sec:Sampling}, we systematically sampled a total of $47,136$ intrinsic binary configurations $\{q,\vec{\chi}_1,\vec{\chi}_2\}$, across the parameter range of the NR surrogate. For the PSD matches, which also require a total mass scale, we choose three masses $M\in\{75,150,250\}M_{\odot}$ compatible with the fixed length of the surrogate waveforms, bringing the total sampled binaries to $141,408$. Full details of the sampling are given in Table~\ref{tab:MatchParams}.

The white match results are shown in Figure~\ref{fig:FullWhiteMatchResults}. We split the results by mass ratio $q\in[1,1.5,2,3,4]$ and mode $(\ell,m)\in[(2\pm2),(2\pm1),(3,\pm3),(4,\pm4)]$. We show the percentages of matches below $0.99$ between the fully spinning waveform, and the waveforms produced by each of the two effective spins, $\nchip$ in orange, and $\chi_p$ in purple. We first note that at the match threshold of $0.99$, we see an improvement by using $\nchip$ over $\chi_p$, across all mass ratios and modes. These improvements are particularly dramatic for higher modes, particularly the $(2,\pm1)$-modes. For example, at mass ratio $q=4$, the percentage of $(2,1)$-mode matches below $0.99$ using the $\chi_p$ parametrization is $96.3\%$, which improves dramatically with $\nchip$ to just $4.3\%$. The parametrizations perform more similarly for the $(2,\pm2)$-modes, but even for the quadrupolar modes we see small improvements. For example, at mass ratio $q=1.5$ we see the percentage of $(2,2)$-mode matches below $0.99$ improving from $8.5\%$ with $\chi_p$ to $4.7\%$ with $\nchip$. 

\begin{figure*}[t!]
\centering
\includegraphics[width=0.8\textwidth]{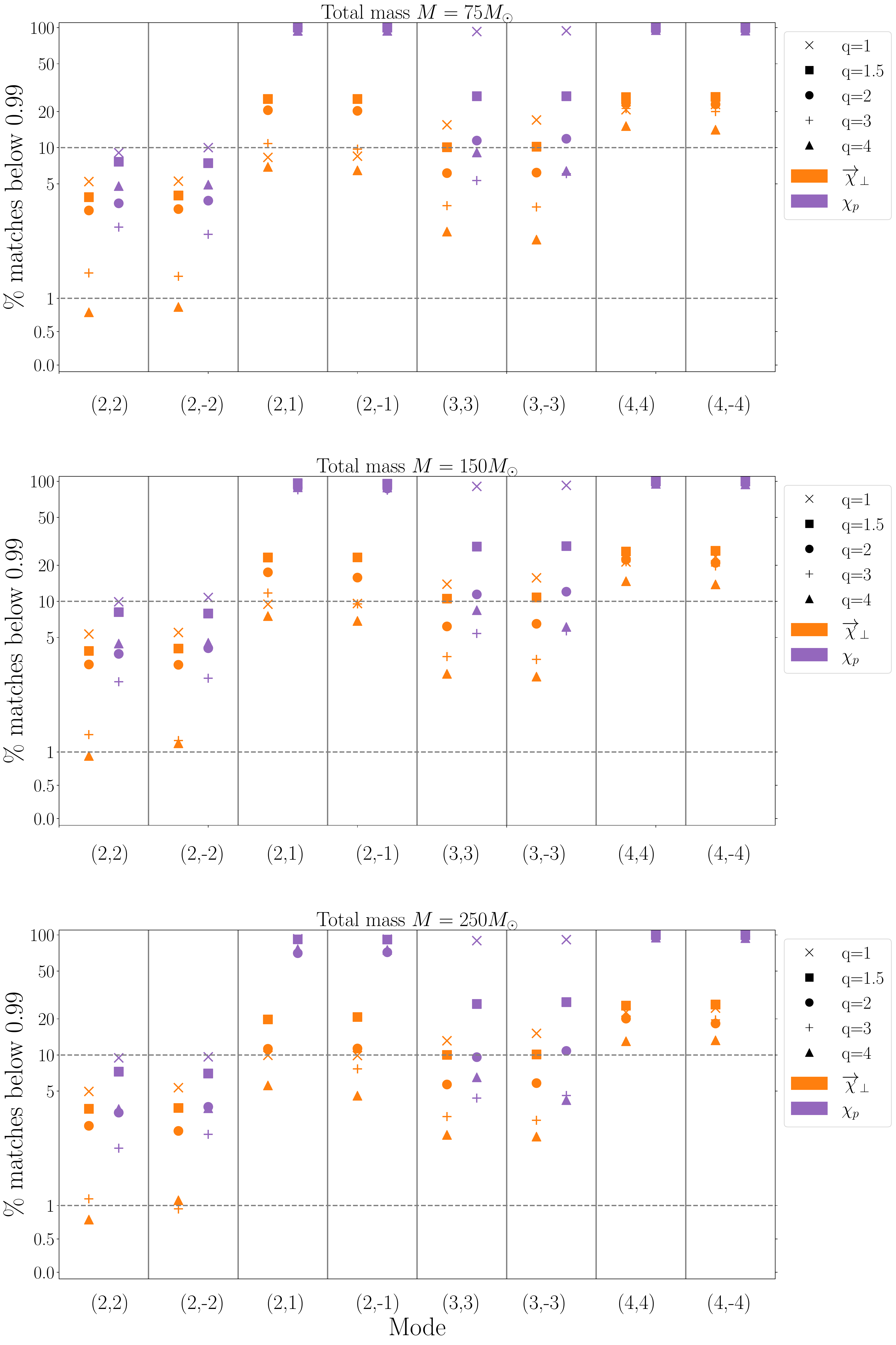}
\caption{Full results for the O4 PSD-weighted matches between the fully spinning waveform and each of the effective spin parametrizations, $\nchip$ (orange) and $\chi_p$ (purple). We show percentages of matches split by total mass, mass ratio, and mode, which have a match less than $0.99$. The dashed horizontal lines indicate $1\%$ and $10\%$. Similar to the white noise matches of Figure~\ref{fig:FullWhiteMatchResults}, we see improvements by using $\nchip$ over $\chi_p$ across all masses, mass ratios and modes, but we note the dramatic improvements in performance for HMs, particularly the $(2,\pm1)$-modes. We also note the high degree of similarity between each of the three total masses.}
\label{fig:FullPSDMatchResults}
\end{figure*}

Using the O4 aLIGO PSD, we similarly split the results by mass ratio and mode, and additionally by the total mass of the system. The results are shown in Figure~\ref{fig:FullPSDMatchResults}. The PSD-weighted match results are qualitatively very similar to the white match results, albeit with small improvements across all matches due to the frequency weighting of the PSD, with only minor differences between each total mass.

\section{Additional Examples}
\label{app:MoreFiducialExamples}
Here we provide additional examples of waveform modes and corresponding precession dynamics produced using the effective spin mappings, both $\chi_p$ and $\nchip$. For each of the three binaries considered here, we show the $(2,1)$-mode, as well as two of the four quaternion component evolutions. In all of the figures, the fully precessing system's results are shown in blue, the results parameterized by $\nchip$ in orange, and by $\chi_p$ in purple. 

\begin{figure*}[ht!]
\centering
\includegraphics[width=\textwidth]{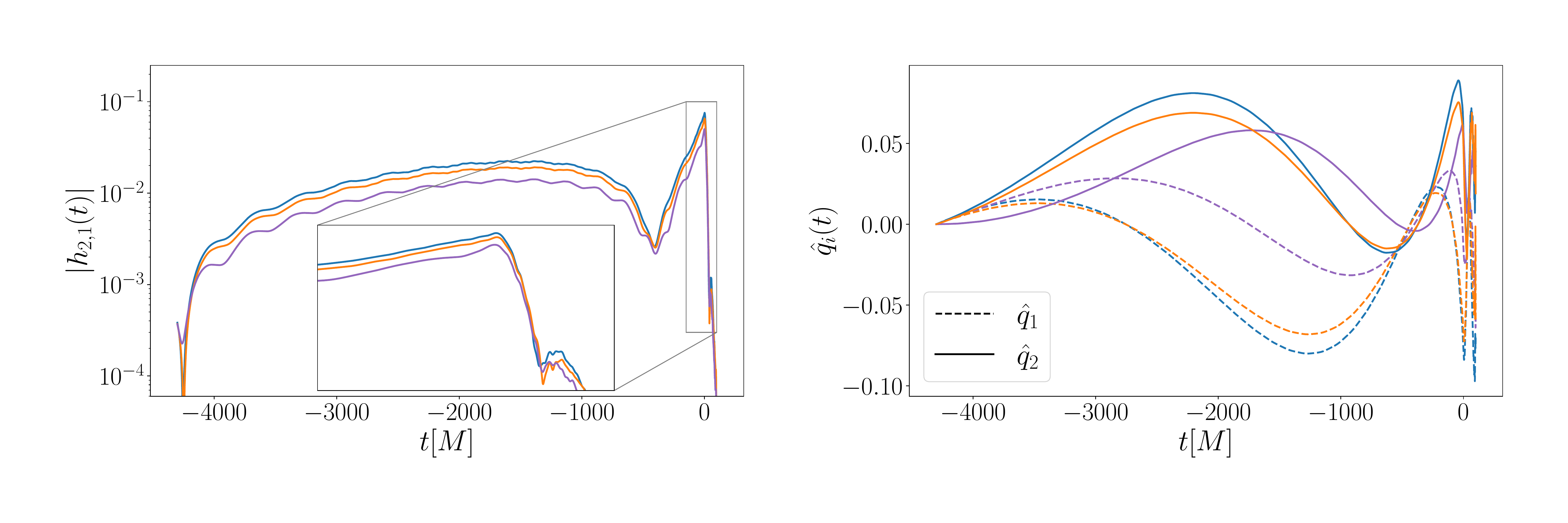}
\caption{Amplitude of the $(2,1)$-mode (left), and time evolution of the two quaternion components, $q_1 (t)$ and $q_2 (t)$ (right), for an equal-mass binary with initial spins $\chi_1(t_0)=(0.225,0.13,-0.15)$, and $\chi_2(t_0)=(0.09,0.15,0.1)$. We show the fully spinning waveform mode and quaternion components in blue. The mode and quaternions parameterized by $\chi_p$ are shown in purple, and $\nchip$ in orange. We see that $\nchip$ more faithfully reproduces the fully precessing $(2,1)$-mode, and much more accurately reproduces the precession dynamics, than $\chi_p$.}
\label{fig:EqualMassFiducial}
\end{figure*}

Our first example is an equal-mass binary, i.e. $q=1$, with initial spins $\vec{\chi}_1(t_0)=(0.225,0.13,-0.15)$, $\vec{\chi}_2(t_0)=(0.09,0.15,0.1)$. In this equal-mass limit, we expect $\chi_p$ to perform poorly, and $\nchip$ to perform much better, as discussed more thoroughly in Sec.~\ref{sec:Discussion}. Indeed, in the left panel of Figure~\ref{fig:EqualMassFiducial}, we see that $\nchip$ better replicates the $(2,1)$-mode for this particular binary, with an amplitude closer to that of the fully precessing waveform and slightly improved phasing. We see the improvement by using $\nchip$ as opposed to $\chi_p$ more clearly in the dynamics as shown in the right panel. The good agreement between the time evolution of fully precessing quaternion components in blue and those of the $\nchip$-mapped system in orange, is in stark contrast to the $\chi_p$-mapped components in purple, which matches the dynamics poorly.

\begin{figure*}[ht!]
\centering
\includegraphics[width=\textwidth]{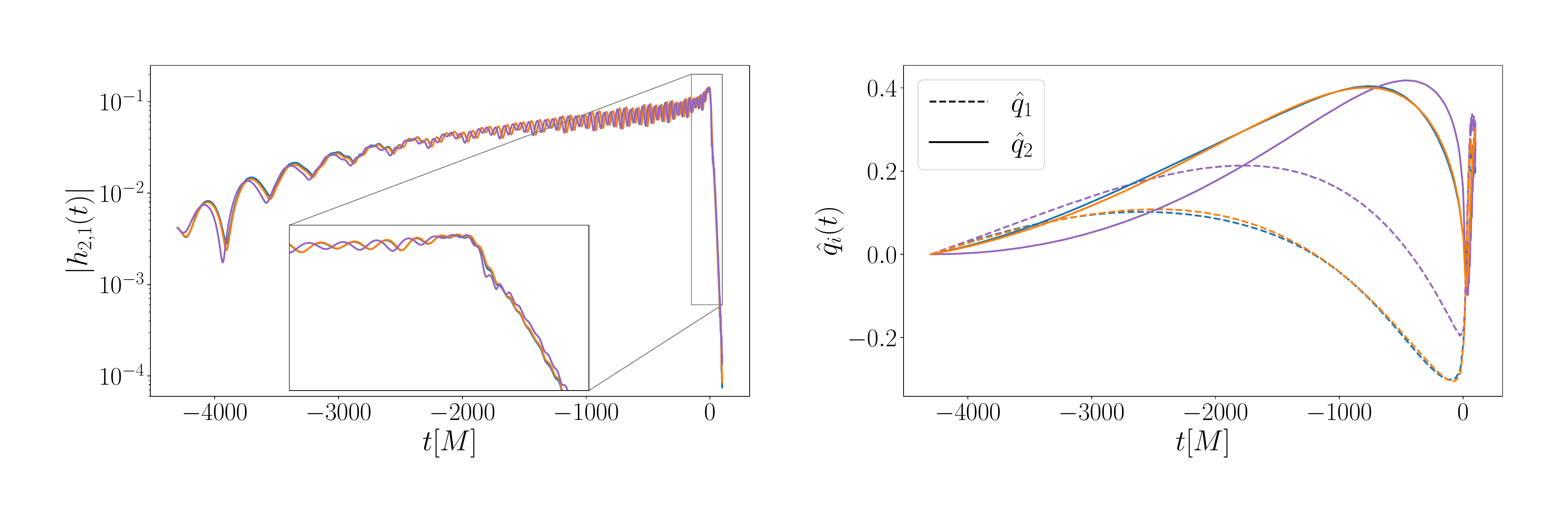}
\caption{Amplitude of the $(2,1)$-mode (left), and time evolution of the two quaternion components, $q_1 (t)$ and $q_2 (t)$ (right), for a $q=3$ fiducial binary, with initial spins $\vec{\chi}_1(t_0)=(0.45,0.26,-0.3)$ and
$\vec{\chi}_2(t_0)=(0.15,0.087,0.1)$. The fully precessing waveform mode and quaternions are shown in blue, with the system parameterized by $\chi_p$ shown in purple, and $\nchip$ in orange. We see that $\nchip$ more faithfully reproduces the fully precessing $(2,1)$-mode, capturing the correct phasing of the mode, and more accurately reproduces the precession dynamics represented by the quaternion components, than $\chi_p$.}
\label{fig:q3Fiducial}
\end{figure*}

The second example is a binary with $q=3$, and initial spins $\vec{\chi}_1(t_0) = (0.45,0.26,-0.3)$, $\vec{\chi}_2(t_0)=(0.15,0.08,0.1)$. We note that in this example, unlike the fiducial binary shown in Figure~\ref{fig:Fiducial}, $\nchip$ is mapped onto the primary BH. The left panel of Figure~\ref{fig:q3Fiducial} shows the $(2,1)$-mode, where we see that unlike in the previous example, the amplitudes of the two mapped waveform modes are very similar to that of the fully precessing mode (blue). However, $\nchip$ clearly much better matches the phasing of the fully precessing mode, with the orange and blue lines being indistinguishable for much of the inspiral, in contrast to $\chi_p$ which shows a clear dephasing, especially in the merger ringdown. We also see that, like for the other fiducial binaries, the quaternion components mapped by $\nchip$, much more faithfully represent the fully precessing quaternions, compared to the $\chi_p$-mapped components, demonstrating that $\nchip$ better replicates the precession dynamics of the fully precessing system.

\begin{figure*}[ht!]
\centering
\includegraphics[width=\textwidth]{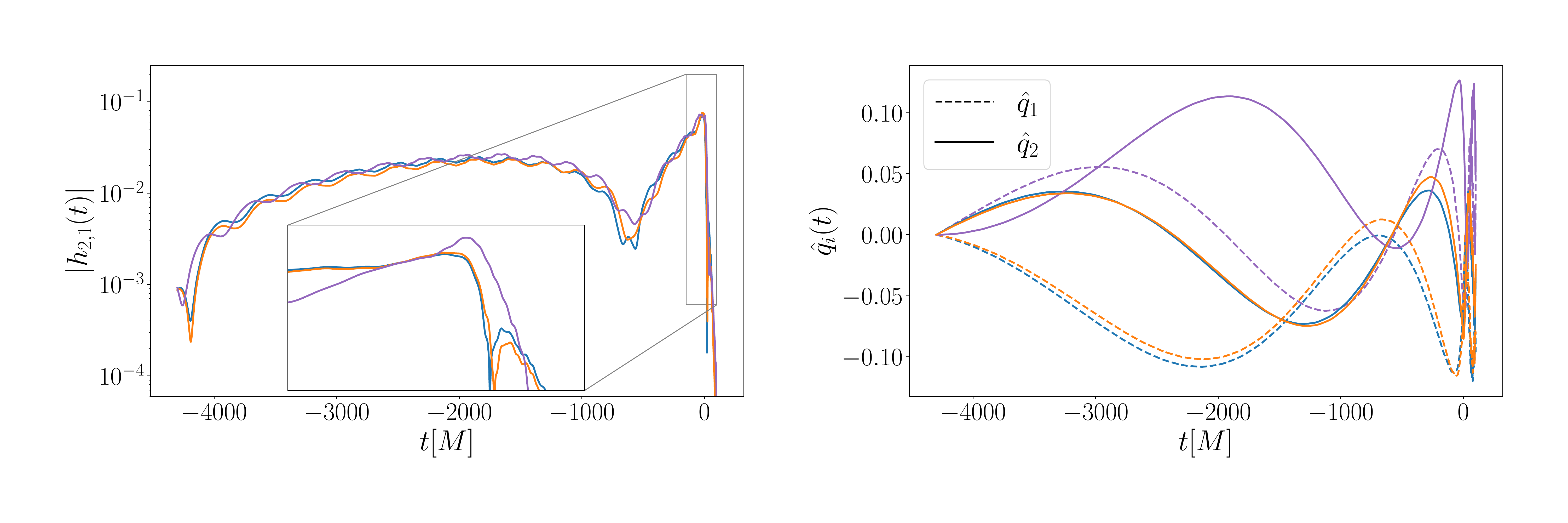}
\caption{Amplitude of the $(2,1)$-mode (left), and time evolution of the two quaternion components, $q_1 (t)$ and $q_2 (t)$ (right), for the same fiducial binary as in Figs.~\ref{fig:Fiducial} and~\ref{fig:quaternions}, but with $\chi_p$ conditionally placed on the secondary BH. The fully precessing waveform mode and quaternions are shown in blue, with the system parameterized by a conditionally placed $\chi_p$ shown in purple, and $\nchip$ in orange. We see that although the conditional placement of $\chi_p$ does lead to an improvement in the accuracy with which it reproduces the $(2,1)$-mode, $\nchip$ still outperforms $\chi_p$. Additionally, conditionally placing $\chi_p$ does not improve the accuracy with which it reproduces the precession dynamics, with $\nchip$ still much more closely matching the time evolution of the fully precessing $\hat{q_1}$ and $\hat{q_2}$ quaternion components.}
\label{fig:FiducialSwitched}
\end{figure*}

Third, we illustrate the impact of the conditional placement 
using the original fiducial binary of Figure~\ref{fig:Fiducial}, but this time showing the effect of an analogous conditional placement with the $\chi_p$-parametrization. In Figure~\ref{fig:FiducialSwitched}, we show the fully precessing fiducial binary in blue, $\nchip$-mapped system in orange, and the $\chi_p$-mapped system with conditional placement in purple. We note that now both $\nchip$ and $\chi_p$ are placed on the secondary BH for this binary. We can see in the left panel that the conditional placement does improve the accuracy with which $\chi_p$ reproduces the $(2,1)$-mode; however the $\nchip$-mapping still reproduces the phasing of the mode significantly better. Additionally, we see in the right panel that the conditional placement of $\chi_p$ does not produce the precession dynamics more accurately, with $\nchip$ still much more closely matching the time evolution of the fully precessing quaternion components.

\begin{figure*}[t!]
\centering
\includegraphics[width=\textwidth]{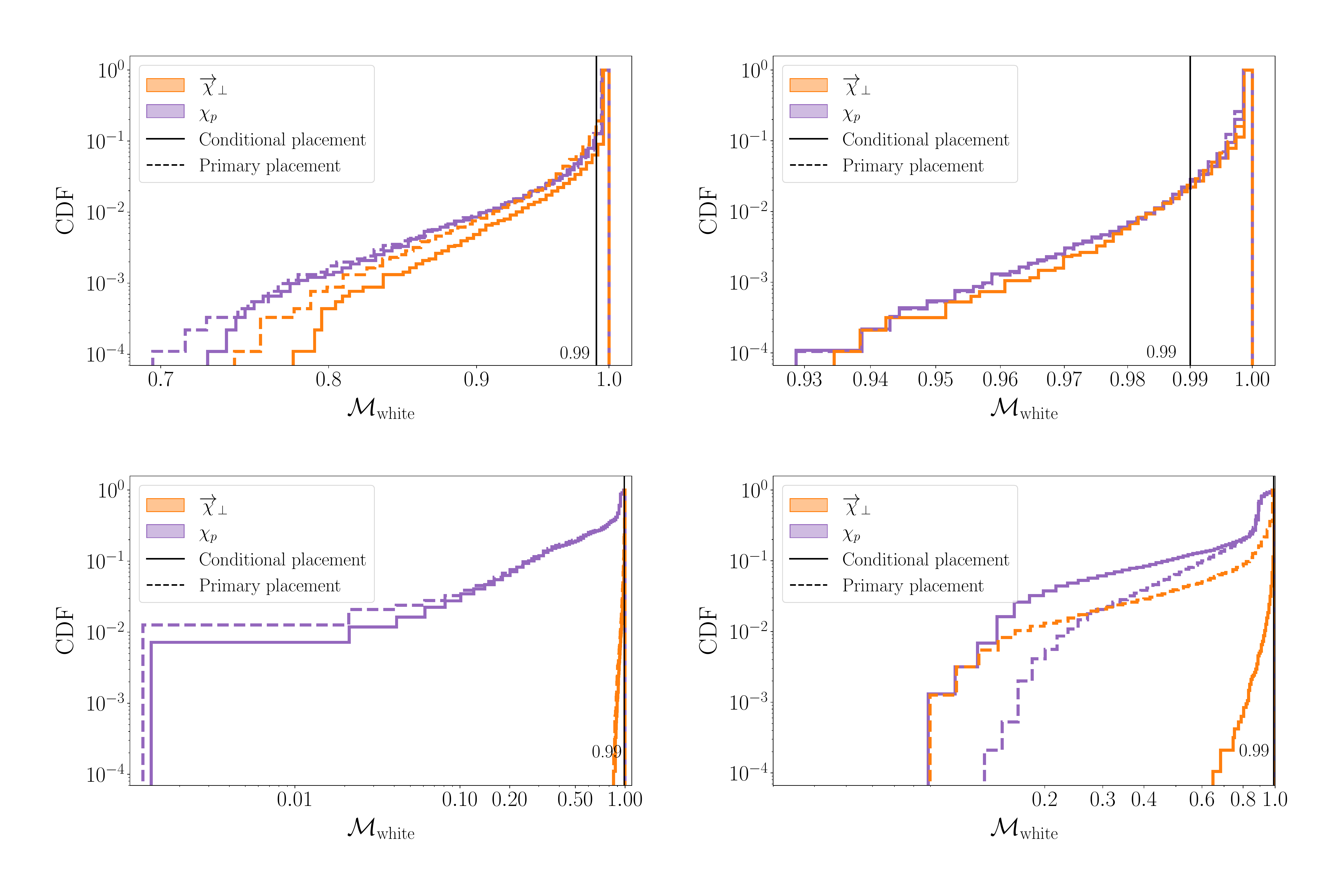}
\caption{Cumulative histograms of white noise mode-by-mode matches for the (2,2)-mode (top row) and the (2,1)-mode (bottom row) for mass ratios $q=1$ (left column) and $q=3$ (right column) for the same binaries as in Figure~\ref{fig:q1whitematchcumul} with the $\nchip$-parametrization (orange) and the $\chi_p$-parametrization (purple). The solid outlines represent the parametrizations including conditional placement, whereas the dashed lines show results when the effective spin is always placed on the primary black hole. The effect of conditional placement is most noticeable at $q=3$ in the (2,1)-mode, where $\nchip$ with conditional placement dramatically outperforms other mappings. We also not that including conditional placement improves the performance of both effective spin parametrizations in most cases, although this improvement is negligible for the (2,2)-mode at mass ratio $q=3$. The exception to this is the $\chi_p$-parametrization for the (2,1)-mode at mass ratio $q=3$, where conditional placement worsens the match distribution for $\chi_p$.}
\label{fig:ExtraWhiteMatches}
\end{figure*}

In addition to these individual cases, in Figure~\ref{fig:ExtraWhiteMatches} we recalculate the white noise mode-by-mode matches shown in Figure~\ref{fig:q1whitematchcumul}, for both spin parametrizations $\chi_p$ and $\nchip$, using (i) conditional placement (solid) and (ii) placement always on the primary black hole (dashed). We note that in all four panels, the best performance is obtained when conditionally placing $\nchip$. The (2,2)-mode at mass ratio $q=1$ (top left) shows a small improvement in both parametrizations' performance when conditional placement is included. The biggest improvement can be seen in the (2,1)-mode at mass ratio $q=3$ (bottom right), where a conditionally placed $\nchip$ dramatically outperforms all other configurations, and neither a conditionally placed $\chi_p$, nor a $\nchip$ affixed to the primary would be able to achieve these improvements. Interestingly, this panel also displays the only instance where conditional placement can worsen the performance of $\chi_p$. Therefore, we conclude that particularly for HMs at unequal-mass ratios, to obtain the dramatic improvements we have seen, both the new effective spin $\nchip$ and conditional placement are required.

\bibliography{References}

\end{document}